\PassOptionsToPackage{table}{xcolor}
\documentclass{article}

\usepackage{worldsonus_preprint,times}
\usepackage{amsmath,amsfonts,bm}

\def\eqref#1{equation~\ref{#1}}

\def\1{\bm{1}}

\DeclareMathAlphabet{\mathsfit}{\encodingdefault}{\sfdefault}{m}{sl}
\SetMathAlphabet{\mathsfit}{bold}{\encodingdefault}{\sfdefault}{bx}{n}

\usepackage{amsmath,amssymb}
\usepackage{booktabs}
\usepackage{array}
\usepackage{tabularx}
\usepackage{caption}
\usepackage{wrapfig}
\usepackage{needspace}
\usepackage{graphicx}
\usepackage{microtype}
\usepackage{multirow}
\usepackage[table]{xcolor}
\definecolor{modelrow}{RGB}{210,230,250}
\usepackage{xspace}
\usepackage{hyperref}
\usepackage{url}
\hypersetup{
  colorlinks=true,
  citecolor=green,
  linkcolor=black,
  linkbordercolor={1 0 0},
  urlcolor=black,
  pdftitle={WorldSonus: Bringing Sound to Worlds},
  pdfauthor={Pengjun Fang, Jingyi Fa, Kam Man Wu, Jiaming Wang, Haoyuan Huang, Yaguang Wu, Xiangjun Huang, Ziyang Ma, Weijia Chen, Hongyu Liu, Zeyue Tian, Qifeng Chen}
}

\usepackage{etoolbox}
\makeatletter
\patchcmd{\find@pdflink}{\Hy@setpdfborder}{%
  \ifnum\pdfstrcmp{#1}{link}=0 /Border[0 0 1]%
  \else\Hy@setpdfborder\fi
}{}{\PackageError{worldsonus}{Unable to configure cross-reference borders}{}}
\makeatother

\newcommand{\modelname}{WorldSonus}
\newcommand{\model}{\modelname\xspace}
\newcommand{\modelours}{\modelname~(Ours)}
\newcommand{\modelheading}{\texorpdfstring{\textnormal{\modelname}}{\modelname}}
\newcommand{\cihalf}[1]{\,{\color{black!50}\pm #1}}

\title{\modelname:   Bringing Sound to Worlds}

\author{%
Pengjun Fang\textsuperscript{1,2}\quad Jingyi Fa\textsuperscript{3}\quad
Kam Man Wu\textsuperscript{1,2}\quad Jiaming Wang\textsuperscript{2}\\[3pt]
\bfseries Haoyuan Huang\textsuperscript{2}\quad Yaguang Wu\textsuperscript{3}\quad
Xiangjun Huang\textsuperscript{3}\quad Ziyang Ma\textsuperscript{4}\\[3pt]
\bfseries Weijia Chen\textsuperscript{2}\quad Hongyu Liu\textsuperscript{1}\quad Zeyue Tian\textsuperscript{\textdagger1,2}\quad
Qifeng Chen\textsuperscript{\textdagger1}\\[7pt]
\textsuperscript{1}The Hong Kong University of Science and Technology\\[2pt]
\textsuperscript{2}Noiz AI\quad\textsuperscript{3}MetaX\\[2pt]
\textsuperscript{4}Shanghai Jiao Tong University%
}

\begin{document}

\maketitle
\begingroup
\renewcommand{\thefootnote}{\fnsymbol{footnote}}
\footnotetext[2]{Corresponding authors.}
\endgroup

\begin{abstract}
Recent advances in world models have enabled increasingly realistic visual synthesis. However, these generated environments remain largely silent. Bringing sound to world models poses three core challenges: \textbf{real-time generation} to keep pace with interactive video streams, \textbf{interactive control} to respond to mid-stream sound instructions, and \textbf{spatially aligned stereo} to reflect scene geometry and camera motion. To address these demands, we introduce \model, an interactive video-to-audio framework designed for real-time spatial sound synthesis in world models. For real-time generation, \model employs a streaming causal autoregressive diffusion architecture that synthesizes audio chunks at a low real-time factor (RTF) of $0.41$. For interactive control, we incorporate an audio-centric captioning pipeline with chunk-indexed prompt scheduling, enabling dynamic manipulation of sound events during generation. For spatial alignment, we leverage high-quality stereo supervision curated from diverse stereo and ambisonic data. Extensive experiments demonstrate that while tailored for world models, \model generalizes effectively to open-domain video-to-audio benchmarks, matching or outperforming state-of-the-art bidirectional models in both acoustic quality and spatial alignment.

Project page: \url{https://noizai.github.io/WorldSonus/}.

\end{abstract}

\section{Introduction}
\label{sec:introduction_hongyu}

Recent advances in video generation~\citep{wan2025,longcatvideo2025,ltxvideo2025,hunyuanvideo2024,cogvideox2024,stepvideo2025} have accelerated the development of generative world models~\citep{gamengen2024,matrixgame22025,worldplay2025,lingbotworld2026}. While these models dynamically synthesize visual environments in response to user actions, the resulting virtual worlds remain predominantly silent. Bringing realistic sound to interactive world models requires audio that seamlessly tracks an evolving visual stream, maintains camera-aligned spatial balance under dynamic viewpoints, and immediately responds to updated text instructions during an ongoing session.

Existing approaches to world-model sound generation fall into two primary categories. The first direction builds upon joint audiovisual foundation models~\citep{ltx22026} extended to long-horizon streaming~\citep{joyaiecho2026,echowm2026,su2026omniforcing}. However, coupling audio synthesis directly to the visual backbone prevents these systems from serving as standalone audio modules for visual-only world models, while causal distillation or long autoregressive (AR) rollouts often accumulate error over time. The second direction employs dedicated video-to-audio (V2A) generators, allowing modular integration with external visual engines. Yet current streaming V2A models remain constrained: V-AURA~\citep{viertola2025vaura} uses bidirectional visual windows and outputs mono audio. SoundReactor~\citep{soundreactor2025} achieves real-time stereo generation but lacks mid-stream text controllability. SwanSphere~\citep{lei2026swansphere} focuses on panoramic video and first-order ambisonics (FOA) soundfields rather than standard perspective streams. Consequently, there remains a need for a modular V2A framework that simultaneously delivers causal real-time generation, mid-stream interactive text control, and spatially aligned perspective stereo audio.

We present \model, a modular V2A framework designed to endow generative world models with real-time, controllable stereo audio. For \textbf{real-time generation}, \model employs a streaming AR-diffusion architecture emitting $100$\,ms audio chunks conditioned only on past and current visual context. A bounded persistent state ensures constant memory usage and compute cost as the stream expands. For \textbf{interactive control}, training on time-varying prompt schedules combined with chunk-level cross-attention enables users to dynamically revise text instructions for on-screen or off-screen sounds without session resets or past-frame recomputation. For \textbf{spatially aligned stereo}, we use clear-stereo clips and convert panoramic FOA data into camera-aligned stereo supervision.

Generating precise audiovisual alignment within causal contexts introduces unique challenges, as future visual frames are unavailable. To enforce temporal synchronization under limited lookahead, we introduce two-timescale visual conditioning and ShiftNCE. Chunk-level visual summaries guide the AR backbone to maintain semantic continuity, while frame-aligned local tokens condition the flow head with fine-grained spatial and temporal cues. To refine event timing without inference cost, ShiftNCE trains the AR state using a contrastive objective supervised by a frozen synchronization expert, eliminating ambiguous temporal negatives during training.

Experiments demonstrate that despite causal operation, \model matches or exceeds state-of-the-art offline bidirectional V2A baselines across both open-domain VGGSound~\citep{chen2020vggsound} videos and our Interactive benchmark, which comprises gameplay and real-world footage. \model achieves VGGish FAD~\citep{hershey2017vggish,kilgour2019fad} scores of $1.73$ on $5$\,s clear-stereo VGGSound~\citep{chen2020vggsound} clips and $2.03$ on $30$\,s interactive videos, with an execution latency of $41.2$\,ms per $100$\,ms audio chunk (real-time factor, $\mathrm{RTF} = 0.41$) on a single NVIDIA H100 GPU.

Our main contributions are threefold:
\begin{itemize}
  \item \textbf{Real-time, interactive stereo generation.}
  We introduce \model, a modular V2A framework that unifies real-time causal generation, mid-stream prompt responsiveness, and camera-aligned stereo synthesis for interactive world models.
  \item \textbf{Causal audiovisual synchronization.}
  We propose two-timescale visual conditioning and a training-only ShiftNCE distillation objective to maintain temporal synchronization under causal streaming without adding runtime latency.
  \item \textbf{State-of-the-art performance.}
  Extensive evaluations show that \model achieves competitive or superior acoustic quality and stereo balance compared with offline bidirectional baselines, with counterfactual tests confirming robust interactive control.
\end{itemize}

\section{Related work}
\label{sec:related}
\paragraph{Bidirectional video-to-audio generation.}
Bidirectional video-to-audio (V2A) synthesis has evolved from text-to-audio latent diffusion models~\citep{liu2023audioldm} and early V2A diffusion networks~\citep{luo2023difffoley,zhang2024foleycrafter} to time-aligned models and rectified flow architectures~\citep{wang2024tiva,mmaudio2024}, improving semantic and temporal fidelity through audio-visual pretraining and explicit synchronization features. Recent models such as AudioX~\citep{tian2025audiox}, ThinkSound~\citep{liu2025thinksound}, and PrismAudio~\citep{liu2026prismaudio} render stereo via flow matching and reward alignment. AC-Foley~\citep{fang2026acfoley} adds reference-audio conditioning, while Omni2Sound~\citep{dai2026omni2sound} investigates video-text-to-audio diffusion across modality combinations. Meanwhile, spatial audio generation has advanced from diffusion-based binaural synthesis~\citep{leng2022binauralgrad} toward object-aware stereo and ambisonics~\citep{karchkhadze2026stereofoley,kim2025visage,liu2025omniaudio}. However, these approaches require full-clip visual context and do not support streaming video input or dynamic control within a persistent causal session.

\paragraph{Streaming video-to-audio generation.}
V-AURA~\citep{viertola2025vaura} uses AR audio-token prediction with chunked waveform decoding, while SwanSphere~\citep{lei2026swansphere} extends causal generation to spatial audio conditioned on panoramic video and text.
Neither establishes fully end-to-end causal streaming: V-AURA conditions each audio step on a bidirectional visual window and non-causal DAC~\citep{kumar2023dac} decoding, and SwanSphere's variational autoencoder (VAE)~\citep{DBLP:journals/corr/KingmaW13}, based on Stable Audio Open~\citep{evans2025stableaudioopen}, does not support stateful stream decoding.
SoundReactor~\citep{soundreactor2025} combines a causal AR backbone with a diffusion head for streaming audio from gameplay video.
Its V2A generator weights are not public, so we adopt only the released
causal VAE.  Building on this line of work, we target open-domain,
interactive video with bounded persistent state and sound instructions that
can change mid-stream.

\paragraph{Interactive world models and joint audio--video generation.}
Interactive video world models such as GameGen-X~\citep{che2025gamegenx} and LingBot-World 2.0~\citep{lingbotworld2026} support controllable visual rollouts and increasingly persistent interaction.  Complementary approaches,
including OmniForcing~\citep{su2026omniforcing} and Ripple~\citep{ding2026ripple}, jointly generate streaming audio and video.  We instead treat audio as a
separate causal module conditioned on an externally produced visual stream:
the upstream model exposes only realized frames, without requiring access to
its action space or internal architecture, while users or agents independently update audio prompts.

\vspace{-0.25em}
\section{\modelheading}
\label{sec:method}
\begin{figure}[t]
  \centering
    \includegraphics[page=1,width=\linewidth]{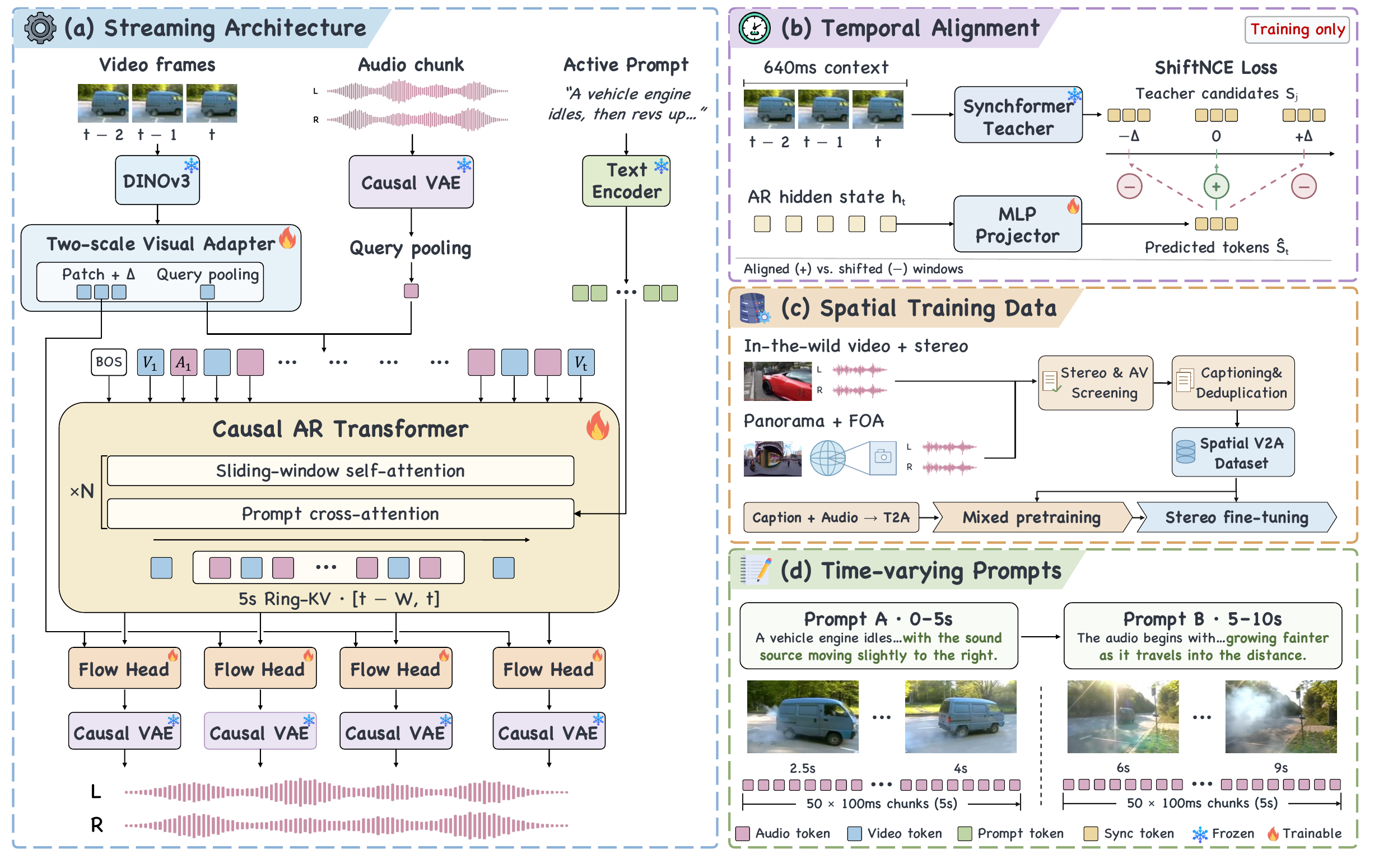}
     \vspace{-1.5em}
    \caption{
    Overview of \model.
    (a) \textbf{Causal streaming pipeline}: Video frames map to two-timescale visual tokens conditioning an AR transformer and a rectified-flow head.
    (b) \textbf{Training-only ShiftNCE}: A training-only frozen Synchformer teacher guides temporal alignment via contrastive window matching.
    (c) \textbf{Spatial stereo supervision}: In-the-wild stereo filtering and panoramic FOA view-decoding provide directional audio grounding.
    (d) \textbf{Interactive prompt control}: Text prompts update dynamically at chunk boundaries, preserving session state and acoustic continuity.}
  \label{fig:overview}
\end{figure}

\vspace{-0.25em}
\model synthesizes real-time stereo audio from streaming video and dynamic prompts via chunk-wise causal AR generation (Figure~\ref{fig:overview}). It couples a bounded recurrent memory with multi-scale visual conditioning and contrastive temporal alignment.
\vspace{-0.25em}
\subsection{Streaming Formulation}
\label{subsec:streaming-formulation}
At each streaming chunk $t \in \{1, \dots, T\}$, the model receives visual frames $\mathbf{v}_t$ and active prompt $\mathbf{p}_t$ to predict the corresponding audio latent chunk $\mathbf{a}_t$. Conditioned on a bounded history state $\mathbf{M}_{t-1}$, the joint distribution factorizes causally over time:
\begin{equation}
    p_\theta(\mathbf{a}_1, \dots, \mathbf{a}_T) = \prod_{t=1}^{T} p_\theta(\mathbf{a}_t \mid \mathbf{M}_{t-1}, \mathbf{v}_t, \mathbf{p}_t),
    \label{eq:streaming-factorization}
\end{equation}
where $\mathbf{M}_{t-1}$ is maintained by a sliding-window Key-Value cache bounded by maximum length $W$.
This factorization uses only frames of past and current chunks, without access to frames of future blocks.
Audio is synthesized in the continuous latent space of a frozen causal stereo VAE~\citep{soundreactor2025}, which encodes $48$\,kHz stereo audio into latents advancing at $30$\,Hz. Its stateful decoder converts predicted chunks to waveform during streaming, with no re-decoding of past latents.
Each $100$\,ms generation chunk corresponds to three latent frames $\mathbf{a}_t \in \mathbb{R}^{3 \times d_a}$, time-aligned with the three visual frames in $\mathbf{v}_t$ ($30$\,FPS).

\vspace{-0.25em}
\subsection{Autoregressive Diffusion}
To balance long-range semantic continuity with high-fidelity continuous generation, we adopt a chunked AR-diffusion framework~\citep{li2024mar,soundreactor2025,lei2026swansphere}. At each step, a decoder-only Transformer processes two chunk-aggregated tokens (one from the previous audio chunk and one from the current video chunk) to maintain session state. Trained with a causal sliding-window attention mask, the model deploys a fixed-capacity Ring-KV cache over context window $W$ (a circular buffer overwriting the oldest slots in place). This ensures bounded memory while matching the context-window policy between training and inference.

The resulting AR hidden state $\mathbf{h}_t$ conditions a compact flow head, which denoises the three latent frames of chunk $t$ using a rectified-flow objective~\citep{lipman2023flow,liu2023rectifiedflow}. The flow head attends bidirectionally inside the current chunk, without access to frames of future blocks, and it reads the frame-aligned visual tokens of Section~\ref{sec:two-timescale} as finer-grained spatial and temporal evidence. Appendix~\ref{app:implementation} gives the remaining network details.
\vspace{-0.25em}
\subsection{Two-Timescale Visual Conditioning}
\label{sec:two-timescale}
To reconcile long-horizon semantic continuity in the bounded AR backbone with precise acoustic rendering in the flow head, we decouple visual conditioning into two complementary timescales: \textbf{chunk-level semantics} and \textbf{frame-aligned local features}. For each chunk (spanning three frames), a frozen DINOv3~\citep{simeoni2025dinov3} encoder maps video frames into spatial patch grids $\mathbf{G}_n$. Mixed-aspect inputs are assigned to resolution buckets and resized without stretching at a fixed pixel budget. To capture motion cues, each grid is concatenated with its temporal difference $\Delta\mathbf{G}_n = \mathbf{G}_n - \mathbf{G}_{n-1}$~\citep{chen2026tdv2a}. A learned query then aggregates each frame's spatial grid into a single feature token, yielding three frame tokens per chunk.

These frame tokens are then processed into the two target timescales:
For \textbf{chunk-level semantics}, a learned summary query aggregates the three frame tokens into a single chunk-level visual token, which is routed to the AR backbone to maintain a compact footprint.
For \textbf{frame-aligned local features}, the three refined frame tokens bypass the AR backbone to condition the flow head directly.
Our dual-path architecture provides the flow head with frame-aligned visual tokens without further chunk-level aggregation, which is essential for preserving the temporal granularity of individual video frames.
\subsection{Interactive Prompt Control}
To support interactive control during streaming, we introduce an in-place prompt control mechanism that allows text instructions to be updated mid-stream.

Given a sound instruction, a T5Gemma~2~\citep{zhang2025t5gemma2} text encoder and compressor project it into a compact token set that conditions the AR backbone via cross-attention, which is cached across streaming steps. When an instruction updates during streaming, the system performs an in-place replacement of the cross-attention cache at the nearest chunk boundary, preserving the Ring-KV cache and existing visual and audio states. To train the model for these dynamic transitions, we incorporate time-varying prompt schedules including mid-sequence instruction switches and drops. This enables seamless steering in a persistent causal session.

\subsection{Training}
\label{sec:training}

\paragraph{Teacher-forced pretraining.}
The generator is pretrained under a rectified flow objective on a mixture of stereo video-audio and audio-only data, using teacher forcing over the bounded context window $W$. For a ground-truth latent chunk $\mathbf{a}_t$, noise $\boldsymbol{\epsilon} \sim \mathcal{N}(\mathbf{0}, \mathbf{I})$, and flow timestep $s \sim \mathcal{U}[0,1]$, the linear interpolation state is $\mathbf{x}_{t,s} = (1-s)\boldsymbol{\epsilon} + s\mathbf{a}_t$. The velocity matching loss is formulated as:
\begin{equation}
    \mathcal{L}_{\mathrm{flow}}(\boldsymbol{\epsilon}) = \left\| f_\theta(\mathbf{x}_{t,s}, s \mid \mathbf{C}_t) - (\mathbf{a}_t - \boldsymbol{\epsilon}) \right\|_2^2,
    \label{eq:flow-objective}
\end{equation}

where $\mathbf{C}_t$ aggregates the AR history state and local frame-aligned visual slots. We apply Explorative Modeling~\citep{gladstone2026explorative} to flow-matching training by evaluating three candidate source noises per step and optimizing the best-matching draw.

\paragraph{Teacher-guided temporal alignment.}
While existing synchronization models~\citep{iashin2024synchformer} provide strong temporal cues, deploying them at inference introduces extra latency and windowing constraints. We instead employ a frozen synchronization encoder as a training-only teacher to distill temporal alignment directly into the generator.

Let $\mathbf{S}_t$ denote the teacher's target embedding for a video window ending at chunk $t$, and let a projector $P(\cdot)$ map the generator's internal AR state $\mathbf{h}_t$ to predict $\mathbf{S}_t$.
To enforce temporal synchronization, our ShiftNCE objective contrasts the aligned target $\mathbf{S}_t$ against temporally shifted embeddings $\mathbf{S}_{t+\delta}$ from the same video across a set of non-zero chunk offsets $\delta \in \mathcal{D}_t$:
\begin{equation}
    \mathcal{L}_{\mathrm{sync}} = -\log \frac{\exp\left( \mathrm{sim}(P(\mathbf{h}_t), \mathbf{S}_t) / \tau \right)}{\exp\left( \mathrm{sim}(P(\mathbf{h}_t), \mathbf{S}_t) / \tau \right) + \sum_{\delta \in \mathcal{D}_t} \exp\left( \mathrm{sim}(P(\mathbf{h}_t), \mathbf{S}_{t+\delta}) / \tau \right)},
    \label{eq:shift-nce}
\end{equation}
where $\mathrm{sim}(\cdot, \cdot)$ denotes cosine similarity, and $\tau$ is the temperature hyperparameter. To prevent false negatives during static visual segments, temporally shifted candidates with high visual similarity under the teacher model are filtered out. The final pretraining objective is the joint loss $\mathcal{L} = \mathcal{L}_{\mathrm{flow}} + \lambda_{\mathrm{sync}}\mathcal{L}_{\mathrm{sync}}$, where $\lambda_{\mathrm{sync}}$ balances the two objectives. Following pretraining, we fine-tune the network on a high-quality stereo video-audio subset. Complete training hyperparameter settings are detailed in Appendix~\ref{app:training-config}.

\section{Data Curation}
\label{sec:data}

\paragraph{Data sources.}
Our dataset contains $1{,}465$ hours of audio, including $999$ hours of paired stereo video-audio data and $466$ hours of audio-only data. The \textit{video-audio subset} uses open-source datasets including VGGSound~\citep{chen2020vggsound}, AudioSet~\citep{gemmeke2017audioset}, Kinetics-700~\citep{carreira2019kinetics700}, and HD-EPIC~\citep{perrett2025hdepic} to cover a wide range of acoustic events. To support audio generation for interactive world models, we collect additional standard stereo video with clear camera motion and dynamic objects. We also incorporate panoramic recordings from the existing Sphere360~\citep{liu2025omniaudio} and YT-AmbiGen~\citep{kim2025visage} datasets. For panoramic clips, we estimate horizontal sound directions from the acoustic energy field to sample perspective visual crops and render view-aligned stereo audio.
The \textit{audio-only subset} includes AudioCaps~\citep{kim2019audiocaps} and isolated audio tracks from weakly-aligned video clips, where the visual content is not aligned with the audio, but the soundtrack retains high-quality acoustic events.

\paragraph{Data filtering.}
To ensure spatial quality and video-audio alignment, we use a two-stage filtering pipeline. First, automated signal processing removes silent, phase-corrupted, narrow-stereo, and dual-mono samples. Second, Qwen3-Omni~\citep{xu2025qwen3omni} acts as a verifier to filter out severe cross-modal mismatches and non-diegetic audio like voiceovers and background music. Complete details appear in Appendix~\ref{app:data-accounting}.

\paragraph{Captioning.}
We generate dense audio-centric captions using Qwen3-Omni~\citep{xu2025qwen3omni}. Prompts focus on sound categories, acoustic texture, temporal timing, and spatial ambiance, while excluding visual details unrelated to sound to avoid introducing irrelevant conditioning cues. Captions are generated from both video and audio for video-audio data, and from audio alone for audio-only data. Annotation details are in Appendix~\ref{app:captioning}.

\section{Experiments}
\label{sec:experiments}

\subsection{Experimental Setup}
\label{sec:setup}

\paragraph{Evaluation sets.}
We evaluate mainly on five test splits: two clear-stereo VGGSound~\citep{chen2020vggsound} splits ($5$\,s and $10$\,s; $4{,}096$ clips each) and three splits comprising interactive gameplay and real-world footage with dynamic camera motion ($5$\,s and $10$\,s, with $4{,}096$ clips each; $30$\,s, with $1{,}024$ clips). We additionally evaluate on physical collision events from Greatest Hits~\citep{owens2016visually}.

\paragraph{Baselines.}
\model is a causal model, using only frames from past and current chunks without access to future frames. In the absence of causal streaming stereo baselines, we evaluate against bidirectional stereo models including AudioX~\citep{tian2025audiox}, ThinkSound~\citep{liu2025thinksound}, and PrismAudio~\citep{liu2026prismaudio}, alongside the streaming monophonic baseline V-AURA~\citep{viertola2025vaura}.

\begin{table}[t]
\caption{Quantitative comparison across five held-out splits ($4{,}096$ clips for $5$\,s/$10$\,s; $1{,}024$ for $30$\,s). Best results are in bold, and second-best are underlined. Acc.: Bi (Bidirectional), Str (Stream), Caus (Causal). Cond.: VT (video+text), V (video-only). Chunk/$t$: chunk length and compute time in ms. IB and CLAP are multiplied by $100$. S-FD$_O$ is omitted for monophonic V-AURA.}
\label{tab:main-results}
\centering
\small
\setlength{\tabcolsep}{2.8pt}
\resizebox{\linewidth}{!}{%
\begin{tabular}{@{}l@{\hspace{4pt}}lccccccccccc@{}}
\toprule
& & & & & \multicolumn{4}{c}{Distribution (mid)} & Spatial & \multicolumn{2}{c}{Semantic} & Temporal \\
\cmidrule(lr){6-9} \cmidrule(lr){10-10} \cmidrule(lr){11-12} \cmidrule(lr){13-13}
Set & Model & Acc. & Cond. & Chunk/$t$ & FAD$\downarrow$ & FD$_P$$\downarrow$ & FD$_O$$\downarrow$ & KL$_P$$\downarrow$ & S-FD$_O$$\downarrow$ & IB$\uparrow$ & CLAP$\uparrow$ & DeSync$\downarrow$ \\
\midrule
\multirow{5}{*}{VGG $5$\,s}
 & AudioX & Bi & VT & -- & 3.00 & 153.29 & \textbf{38.19} & \underline{1.66} & 87.60 & \underline{26.96} & 37.87 & 0.986 \\
 & ThinkSound & Bi & VT & -- & 2.94 & \underline{141.40} & 47.32 & 1.91 & \underline{62.95} & 24.12 & 33.62 & \textbf{0.481} \\
 & PrismAudio & Bi & VT & -- & \underline{2.09} & \textbf{132.58} & 50.72 & 1.67 & 94.77 & 26.28 & \textbf{39.20} & \underline{0.539} \\
 & V-AURA & Str & V & $640$/$636$ & 4.06 & 304.06 & 51.01 & 2.03 & -- & 26.48 & 24.91 & 1.287 \\
\rowcolor{modelrow} & \modelours & Caus & VT & $100$/$41.2$ & \textbf{1.73} & 159.74 & \underline{39.03} & \textbf{1.48} & \textbf{43.47} & \textbf{27.06} & \underline{38.35} & 0.686 \\
\midrule
\multirow{5}{*}{VGG $10$\,s}
 & AudioX & Bi & VT & -- & 3.86 & 194.72 & \textbf{41.50} & 1.54 & 83.53 & 27.84 & \underline{34.56} & 0.944 \\
 & ThinkSound & Bi & VT & -- & \underline{2.18} & \textbf{144.17} & 46.49 & 1.73 & \underline{55.02} & 25.76 & 30.46 & \underline{0.514} \\
 & PrismAudio & Bi & VT & -- & 2.64 & \underline{154.69} & 52.38 & \underline{1.47} & 118.08 & \underline{28.16} & 33.26 & \textbf{0.427} \\
 & V-AURA & Str & V & $640$/$636$ & 4.35 & 318.13 & 57.71 & 1.95 & -- & 26.73 & 23.35 & 1.208 \\
\rowcolor{modelrow} & \modelours & Caus & VT & $100$/$41.2$ & \textbf{1.79} & 167.52 & \underline{41.90} & \textbf{1.32} & \textbf{45.40} & \textbf{28.82} & \textbf{36.77} & 0.687 \\
\midrule
\multirow{5}{*}{Inter.\ $5$\,s}
 & AudioX & Bi & VT & -- & \underline{4.44} & \underline{186.49} & 66.77 & \underline{1.56} & 109.21 & 24.88 & \underline{32.53} & 1.059 \\
 & ThinkSound & Bi & VT & -- & 8.49 & 256.38 & \underline{61.84} & 1.98 & \underline{58.68} & 18.39 & 22.15 & \underline{0.733} \\
 & PrismAudio & Bi & VT & -- & 6.22 & 222.83 & 62.74 & 1.60 & 83.85 & 23.08 & 32.42 & \textbf{0.651} \\
 & V-AURA & Str & V & $640$/$636$ & 8.18 & 436.10 & 83.78 & 2.01 & -- & \textbf{26.84} & 16.07 & 1.344 \\
\rowcolor{modelrow} & \modelours & Caus & VT & $100$/$41.2$ & \textbf{2.68} & \textbf{163.74} & \textbf{40.99} & \textbf{1.29} & \textbf{34.26} & \underline{26.48} & \textbf{35.90} & 0.831 \\
\midrule
\multirow{5}{*}{Inter.\ $10$\,s}
 & AudioX & Bi & VT & -- & \underline{5.05} & 352.02 & 69.19 & 1.57 & 119.16 & 25.39 & 27.89 & 1.077 \\
 & ThinkSound & Bi & VT & -- & 6.27 & \underline{332.49} & \underline{60.93} & 1.85 & \underline{56.47} & 21.43 & 24.10 & \underline{0.738} \\
 & PrismAudio & Bi & VT & -- & 5.16 & 338.17 & 61.23 & \underline{1.42} & 96.12 & 25.20 & \underline{31.08} & \textbf{0.599} \\
 & V-AURA & Str & V & $640$/$636$ & 9.75 & 532.74 & 86.46 & 1.96 & -- & \underline{27.60} & 15.47 & 1.192 \\
\rowcolor{modelrow} & \modelours & Caus & VT & $100$/$41.2$ & \textbf{2.62} & \textbf{262.25} & \textbf{42.91} & \textbf{1.30} & \textbf{37.42} & \textbf{28.63} & \textbf{34.22} & 0.820 \\
\midrule
\multirow{5}{*}{Inter.\ $30$\,s}
 & AudioX & Bi & VT & -- & \underline{5.52} & 427.00 & 81.14 & 1.88 & 121.62 & \textbf{24.78} & 30.34 & 1.125 \\
 & ThinkSound & Bi & VT & -- & 7.00 & 473.40 & 73.46 & 2.31 & \underline{81.88} & 11.71 & 19.50 & 0.901 \\
 & PrismAudio & Bi & VT & -- & 6.08 & \underline{375.22} & \underline{62.99} & \underline{1.66} & 95.26 & 22.32 & \underline{33.70} & \textbf{0.765} \\
 & V-AURA & Str & V & $640$/$636$ & 13.33 & 666.31 & 108.01 & 2.74 & -- & 19.93 & 8.54 & 1.266 \\
\rowcolor{modelrow} & \modelours & Caus & VT & $100$/$41.2$ & \textbf{2.03} & \textbf{253.21} & \textbf{26.67} & \textbf{1.22} & \textbf{25.44} & \underline{22.93} & \textbf{34.21} & \underline{0.867} \\
\bottomrule
\vspace{-2em}
\end{tabular}}
\end{table}

\subsection{Metrics}
\label{sec:eval-metrics}
\paragraph{Audio quality and audiovisual alignment.}
For \textit{acoustic fidelity}, we report Fr\'echet distances in VGGish~\citep{hershey2017vggish} (FAD~\citep{kilgour2019fad}), PaSST~\citep{koutini2022passt} (FD$_P$), and OpenL3~\citep{cramer2019openl3} (FD$_O$) feature spaces, together with PaSST-based KL divergence (KL$_P$). FAD and FD$_O$ use the mid-channel signal. S-FD$_O$ applies the OpenL3 distance to the side-channel signal to evaluate stereo distribution quality. For \textit{audiovisual alignment}, we use ImageBind~\citep{girdhar2023imagebind} (IB) similarity for semantic consistency and Synchformer~\citep{iashin2024synchformer} offset (DeSync) for synchronization, supplemented by onset accuracy, F1, and AP on Greatest Hits following MMAudio~\citep{mmaudio2024}.

\paragraph{Stereo balance agreement.}
BiasSkill measures whether generated and reference audio favor the same left/right channel within $1$\,s windows. It reports the normalized gain over a permutation baseline that accounts for each model's fixed channel bias. Details are provided in Appendix~\ref{app:biasskill}.

\subsection{Main Results}
\label{sec:main-results}
\paragraph{Audio quality.}
We conduct quantitative comparisons across all five held-out test splits. As shown in Table~\ref{tab:main-results}, \model achieves competitive or superior acoustic quality (FAD, FD$_P$, and KL$_P$) compared to bidirectional models on both VGGSound and interactive video datasets, despite using only visual frames from past and current chunks. Compared with the streaming baseline V-AURA, \model achieves lower distributional divergence and stronger audiovisual alignment while running at a much finer temporal resolution ($100$\,ms vs.\ $640$\,ms), synthesized in just $41.2$\,ms per chunk on a single NVIDIA H100 GPU ($\mathrm{RTF}=0.41$) to comfortably satisfy real-time streaming requirements. Together, these results demonstrate that bounded-memory causal generation can combine competitive audio quality with real-time synthesis across the evaluated $5$--$30$\,s durations.

\begin{figure}[h]
\centering
\includegraphics[width=\linewidth]{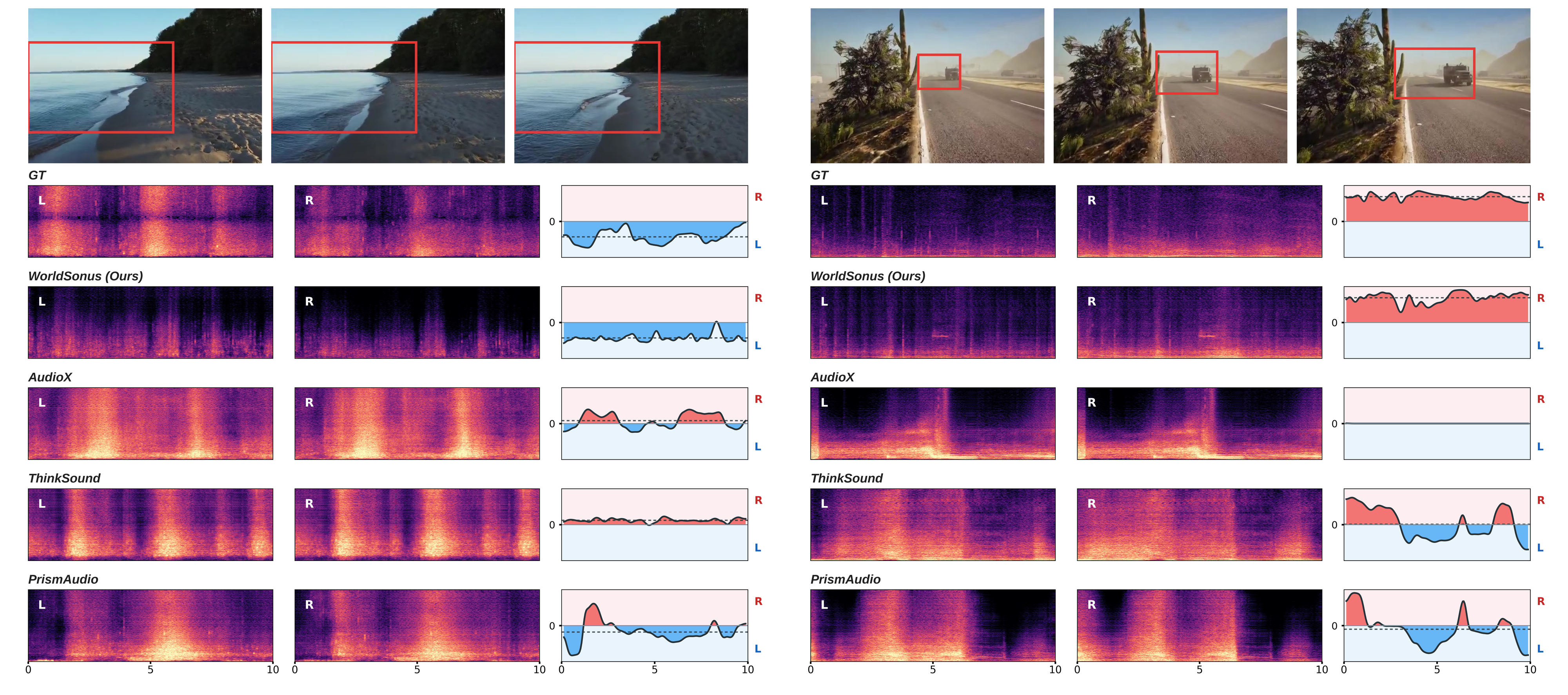}
\caption{Qualitative stereo layout versus bidirectional baselines on two interactive clips. Red boxes mark sources. Energy-balance curves show left (negative) versus right (positive) channel dominance.}
\label{fig:spatial-qual}
\vspace{-8pt}
\end{figure}

\paragraph{Spatial and temporal alignment.}
\Needspace{16\baselineskip}
\begin{wraptable}{r}{0.52\linewidth}
\vspace{-1.5em}
\captionsetup{position=top}
\captionof{table}{Stereo balance agreement (BiasSkill, \%). Full table in App.~\ref{app:biasskill}.}
 \vspace{-0.8em}
\label{tab:biasskill}
\small
\setlength{\tabcolsep}{4.5pt}
\begin{tabular}{@{}lccc@{}}
\toprule
Model & VGG $10$\,s & Inter.\ $10$\,s & Inter.\ $30$\,s \\
\midrule
AudioX & \underline{3.79} & \underline{3.08} & ${-}0.01$ \\
ThinkSound & 3.22 & ${-}1.78$ & 0.63 \\
PrismAudio & 1.10 & ${-}0.06$ & \underline{4.19} \\
\multicolumn{1}{@{}>{\columncolor{modelrow}[0pt][\tabcolsep]}l}{\modelours}
& \cellcolor{modelrow}\textbf{4.42} & \cellcolor{modelrow}\textbf{6.03}
& \multicolumn{1}{>{\columncolor{modelrow}[\tabcolsep][0pt]}c@{}}{\textbf{15.90}} \\
\bottomrule
\end{tabular}

\vspace{0.8em}

\centering
\captionof{table}{Onset detection on Greatest Hits.}
\vspace{-0.8em}
\label{tab:onset-gh}
\small
\setlength{\tabcolsep}{12.4pt}
\begin{tabular}{@{}lccc@{}}
\toprule
Model & Acc.$\uparrow$ & F1$\uparrow$ & AP$\uparrow$ \\
\midrule
AudioX & \textbf{0.858} & 0.770 & 0.790 \\
ThinkSound & 0.706 & 0.758 & \underline{0.904} \\
PrismAudio & 0.783 & \textbf{0.808} & \textbf{0.906} \\
w/o finetune & \underline{0.805} & \underline{0.796} & 0.884 \\
w/o ShiftNCE & 0.678 & 0.560 & 0.681 \\
\multicolumn{1}{@{}>{\columncolor{modelrow}[0pt][\tabcolsep]}l}{\modelours}
& \cellcolor{modelrow}0.803 & \cellcolor{modelrow}0.785
& \multicolumn{1}{>{\columncolor{modelrow}[\tabcolsep][0pt]}c@{}}{0.871} \\
\bottomrule
\end{tabular}
\vspace{-2em}
\end{wraptable}
For spatial alignment, we complement distributional stereo metrics (S-FD$_O$ in Table~\ref{tab:main-results}) with stereo balance agreement measured by BiasSkill over $1$\,s windows (Table~\ref{tab:biasskill}).
\model achieves the highest BiasSkill across all evaluated splits.
Figure~\ref{fig:spatial-qual} provides complementary qualitative evidence from two interactive clips with sound sources located predominantly on one side: beach surf on the left and a passing truck on the right.
The left/right spectrograms and signed energy-balance curves $(E_R - E_L)/(E_R + E_L)$ show that \model reproduces the corresponding channel dominance, consistent with the ground-truth audio, whereas the baselines exhibit weaker or less consistent spatial alignment.

For temporal alignment, bidirectional baselines that optimize directly on Synchformer features achieve lower DeSync (Table~\ref{tab:main-results}).
Nevertheless, \model maintains competitive event timing without test-time guidance, as reflected in Table~\ref{tab:onset-gh} on the model-free Greatest Hits onset benchmark.

Together, these results show that \model combines competitive temporal alignment with stronger spatial alignment under causal visual conditioning.

\Needspace{6\baselineskip}
\paragraph{Long-horizon stability.}
To analyze potential drift during continuous streaming, we evaluate $1{,}024$ Interactive $30$\,s clips under two regimes: \textit{Direct} (generating the final $25$--$30$\,s window from a cold start) and \textit{Rollout tail} (extracting the final window from an uninterrupted $30$\,s stream). As shown in Table~\ref{tab:long-horizon}, the rollout tail closely tracks the direct baseline in acoustic fidelity and synchronization (FAD $2.51$ vs.\ $2.63$, DeSync $0.827$ vs.\ $0.880$), with minimal shifts in OpenL3 and ImageBind. It indicates that under our bounded $5$\,s Ring-KV cache, \model maintains stable, collapse-free streaming across extended horizons without periodic resets.

\begin{table}[!ht]
\caption{Long-horizon stability on Interactive $30$\,s ($1{,}024$ clips). Final $5$\,s window ($25$--$30$\,s) generated directly versus via continuous rollout.}
\label{tab:long-horizon}
\centering
\small
\setlength{\tabcolsep}{5pt}
\begin{tabular*}{\linewidth}{@{\extracolsep{\fill}}lccccccc@{}}
\toprule
Protocol & FAD$\downarrow$ & FD$_P$$\downarrow$ & FD$_O$$\downarrow$ & KL$_P$$\downarrow$ & S-FD$_O$$\downarrow$ & IB$\uparrow$ & DeSync$\downarrow$ \\
\midrule
Direct last $5$\,s & 2.63 & 271.68 & \textbf{35.16} & \textbf{1.22} & \textbf{39.24} & \textbf{21.38} & 0.880 \\
Rollout tail ($25$--$30$\,s) & \textbf{2.51} & \textbf{246.69} & 40.46 & 1.31 & 41.77 & 19.99 & \textbf{0.827} \\
\bottomrule
\end{tabular*}
\end{table}

\Needspace{23\baselineskip}
\begin{wraptable}{r}{0.5\linewidth}
\vspace{-0.6em}
\centering
\captionsetup{position=top,skip=5pt}
\caption{Dual relative match rate (\%). Both halves prefer their own reference prompt.}
\label{tab:prompt-matching-baselines}
\small
\setlength{\tabcolsep}{7pt}
\begin{tabularx}{\linewidth}{@{}Xcc@{}}
\toprule
Model & VGG $10$\,s & Inter.\ $10$\,s \\
\midrule
AudioX & 15.41 & \underline{14.04} \\
ThinkSound & 13.92 & 13.43 \\
PrismAudio & \underline{16.16} & 11.99 \\
V-AURA (video-only) & 12.16 & 12.11 \\
\multicolumn{1}{@{}>{\columncolor{modelrow}[0pt][\tabcolsep]}l}{\modelours}
 & \cellcolor{modelrow}\textbf{25.68}
 & \multicolumn{1}{>{\columncolor{modelrow}[\tabcolsep][0pt]}c@{}}{\textbf{23.44}} \\
\bottomrule
\end{tabularx}
\vspace{1.2em}
\caption{Paired prompt-switch control. Gain $G$ is switch-minus-hold preference. $\pm$ is half the $95\%$ percentile bootstrap interval (App.~\ref{app:prompt-control}).}
\label{tab:prompt-control}
\small
\setlength{\tabcolsep}{4.5pt}
\begin{tabular*}{\linewidth}{@{\extracolsep{\fill}}lccc@{}}
\toprule
Set & Switch pref. & Hold pref. & Gain $G$ \\
\midrule
VGG $10$\,s & $+0.024$ & $-0.027$ & $0.051\cihalf{0.004}$ \\
Inter.\ $10$\,s & $+0.022$ & $-0.030$ & $0.053\cihalf{0.003}$ \\
\bottomrule
\end{tabular*}
\vspace{-1.5em}
\end{wraptable}

\paragraph{Text control.}
\label{sec:prompt-control}
To evaluate interactive prompt control, we measure text to audio agreement using CLAP~\citep{wu2023clap} between each $5$\,s audio segment and its caption. For $10$\,s clips, let $P_1$ and $P_2$ denote the sequential instructions for the first and second $5$\,s halves ($P_1 \to P_2$). We report the dual relative match rate, defined as the percentage of clips where both $5$\,s segments score higher with their own caption than with the other. As shown in Table~\ref{tab:prompt-matching-baselines}, \model achieves dual match rates of $25.68\%$ on VGGSound and $23.44\%$ on Interactive $10$\,s, outperforming all baseline architectures.

However, strong text-audio alignment alone does not fully isolate text responsiveness from natural visual transitions. To rule out visual conflation, we perform paired counterfactual interventions by comparing a Switch setting ($P_1 \to P_2$ at $5$\,s) against a Hold baseline ($P_1 \to P_1$) under identical visual features and seeds. The intervention gain $G$ measures how much switching increases the second half's CLAP preference for $P_2$ over $P_1$. As shown in Table~\ref{tab:prompt-control}, \model yields positive net intervention gains $G$ ($+0.0509$ and $+0.0525$) with $95\%$ bootstrap confidence intervals excluding zero. These combined results confirm that \model achieves active, causal text steering over generated audio with visual context held fixed. Details are provided in Appendix~\ref{app:prompt-control}.

\begin{figure}[h]
\centering
\includegraphics[width=\linewidth]{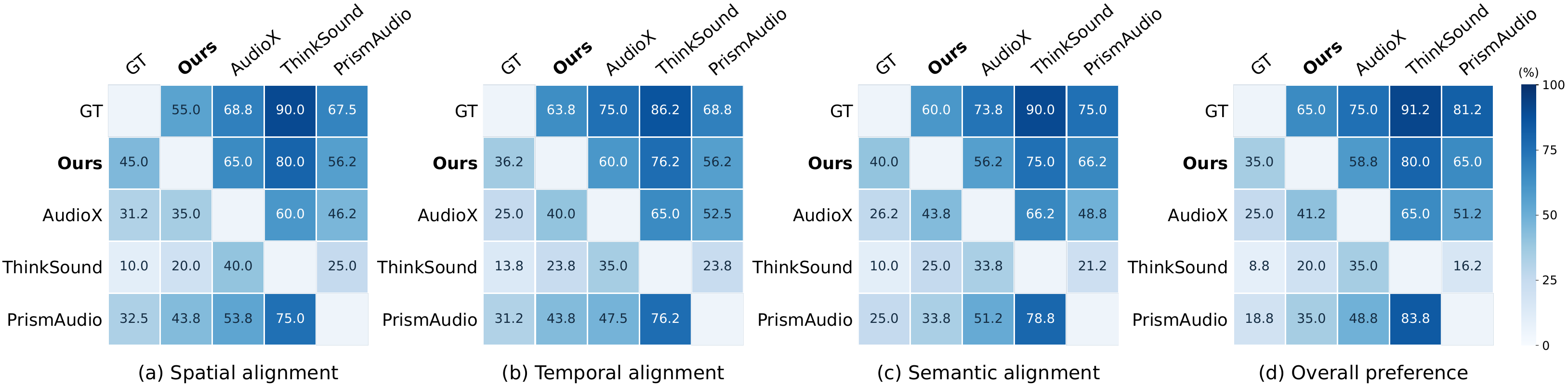}
\caption{User-study preference matrices across (a) spatial alignment, (b) temporal alignment, (c) semantic alignment, and (d) overall preference. Cell $(i,j)$ is the percentage preferring the row method over the column method, counting each tie as half a vote ($40$ ratings per pair).}
\label{fig:user-study}
\vspace{-4pt}
\end{figure}

\vspace{-0.5em}
\paragraph{Subjective comparison.}
To complement objective evaluations, we perform a blind, randomized human A/B study on $40$ representative Interactive $10$\,s clips. Twenty assessors provided $400$ pairwise evaluations across four criteria: spatial alignment, temporal alignment, semantic match, and overall preference, with ties counted as half a vote (Appendix~\ref{app:user-study}). As shown in Figure~\ref{fig:user-study}, \model is preferred over all three generated baselines across all criteria, while ground-truth audio remains preferred. Overall preference reaches $58.8\%$ against AudioX, $80.0\%$ against ThinkSound, and $65.0\%$ against PrismAudio, indicating stronger overall listener preference.

\vspace{-0.5em}

\subsection{Ablation Studies}
\label{subsec:ablations}
All ablation checkpoints are evaluated directly after pretraining without fine-tuning on the Interactive 10\,s set ($4{,}096$ clips) under fixed classifier-free guidance~\citep{ho2022cfg}. Table~\ref{tab:ablations} isolates architectural and alignment components, while Table~\ref{tab:chunk-ablation} compares generation-chunk granularities.

\textbf{Temporal alignment objective.}
To investigate whether the proposed ShiftNCE objective explicitly drives fine-grained synchronization, and whether external temporal representations can replace it, we compare our default model against a variant trained without ShiftNCE and another incorporating direct features from a $640$\,ms causal Synchformer~\citep{iashin2024synchformer} window. As shown in Table~\ref{tab:ablations}, removing ShiftNCE degrades DeSync from $0.831$ to $1.067$, while direct feature injection yields worse timing and acoustic quality, possibly reflecting a mismatch between its $640$\,ms feature window and our $100$\,ms streaming chunks. These results confirm that ShiftNCE provides the necessary temporal supervision to sharpen alignment without requiring additional sync encoders or inference overhead.

\Needspace{21\baselineskip}
\begin{wraptable}{r}{0.55\linewidth}
\vspace{-1.2em}
\captionsetup{position=top,skip=4pt}
\centering
\captionof{table}{Architecture ablations on Interactive $10$\,s.}
\label{tab:ablations}
\fontsize{6.5}{8}\selectfont
\setlength{\tabcolsep}{1.2pt}

\begin{tabularx}{\linewidth}{@{}p{0.27\linewidth}*{4}{>{\hsize=0.91\hsize\centering\arraybackslash}X}>{\hsize=1.25\hsize\centering\arraybackslash}X>{\hsize=0.91\hsize\centering\arraybackslash}X>{\hsize=1.2\hsize\centering\arraybackslash}X@{}}
\toprule
Variant & FAD$\downarrow$ & FD$_P$$\downarrow$ & FD$_O$$\downarrow$ & KL$_P$$\downarrow$ & S-FD$_O$$\downarrow$ & IB$\uparrow$ & DeSync$\downarrow$ \\
\midrule
w/o ShiftNCE & \underline{2.49} & 276.68 & 48.50 & 1.44 & 54.08 & 27.20 & 1.067 \\
Direct sync features & 2.81 & 278.28 & 46.33 & 1.40 & 48.68 & 26.22 & 1.043 \\
w/o temporal delta & 3.08 & \underline{267.51} & 49.37 & 1.40 & 68.76 & \textbf{28.77} & 0.958 \\
Pooled SigLIP~2 & 3.29 & 291.60 & \textbf{31.96} & \underline{1.33} & \textbf{29.81} & 24.56 & \underline{0.944} \\
w/o flow-head DINO & 15.82 & 906.49 & 117.12 & 2.25 & 127.59 & 12.65 & 1.020 \\
\midrule
\rowcolor{modelrow} \modelours & \textbf{2.42} & \textbf{266.41} & \underline{40.97} & \textbf{1.31} & \underline{40.17} & \underline{28.05} & \textbf{0.831} \\
\bottomrule
\end{tabularx}

\vspace{1.5em}

\centering
\captionof{table}{Chunk-length comparison on Interactive $10$\,s.}
\label{tab:chunk-ablation}
\fontsize{6.5}{8}\selectfont
\setlength{\tabcolsep}{1.2pt}

\begin{tabularx}{\linewidth}{@{}p{0.27\linewidth}*{4}{>{\hsize=0.91\hsize\centering\arraybackslash}X}>{\hsize=1.25\hsize\centering\arraybackslash}X>{\hsize=0.91\hsize\centering\arraybackslash}X>{\hsize=1.2\hsize\centering\arraybackslash}X@{}}
\toprule
Chunk & FAD$\downarrow$ & FD$_P$$\downarrow$ & FD$_O$$\downarrow$ & KL$_P$$\downarrow$ & S-FD$_O$$\downarrow$ & IB$\uparrow$ & DeSync$\downarrow$ \\
\midrule
$33$\,ms & 3.79 & 333.05 & 62.17 & 1.68 & 62.16 & \underline{27.18} & 0.915 \\
$200$\,ms & \underline{2.51} & \underline{274.86} & \textbf{39.03} & \textbf{1.26} & \textbf{35.12} & 26.35 & \textbf{0.819} \\
\midrule
\rowcolor{modelrow} $100$\,ms & \textbf{2.42} & \textbf{266.41} & \underline{40.97} & \underline{1.31} & \underline{40.17} & \textbf{28.05} & \underline{0.831} \\
\bottomrule
\end{tabularx}
\vspace{-2em}
\end{wraptable}

\textbf{Visual feature representation.}
To examine whether our DINOv3 patch-based representation outperforms the pooled CLIP-family features commonly used in prior V2A models~\citep{zhang2024foleycrafter,mmaudio2024,liu2025thinksound,shan2025hunyuanfoley}, we compare it with pooled SigLIP~2~\citep{tschannen2025siglip2} features. We also evaluate a variant without temporal deltas ($\Delta\mathbf{G}_n$). Table~\ref{tab:ablations} shows that our DINOv3 features substantially improve ImageBind alignment from $24.56$ to $28.05$ over the Pooled SigLIP~2 baseline. Furthermore, removing temporal deltas worsens both DeSync ($0.958$) and FAD ($3.08$). These results demonstrate the effectiveness of our DINOv3-based visual conditioning and explicit temporal cues for video-to-audio synthesis.

\textbf{Two-timescale visual conditioning.}
To evaluate the role of feeding direct frame-level visual features to the flow head, we compare our model with a variant that removes this shortcut path and passes visual information only through the autoregressive state. As shown in Table~\ref{tab:ablations}, removing the direct visual connection leads to a severe degradation in audio quality, with FAD increasing from $2.42$ to $15.82$. This confirms that in our streaming design, combining direct frame-level visual inputs with higher-level autoregressive context is essential for generating clear and synchronized audio.

\textbf{Generation-chunk granularity.}
To determine the optimal balance between streaming latency and generation quality, we evaluate chunk sizes across $33$\,ms, $100$\,ms, and $200$\,ms. Table~\ref{tab:chunk-ablation} shows that shrinking the chunk size to $33$\,ms worsens acoustic quality (FAD $3.79$), whereas increasing it to $200$\,ms increases latency without consistent improvements across metrics. We therefore choose $100$\,ms as a practical balance between streaming granularity and generation quality.

\section{Conclusion}
We present \model, a causal spatial audio synthesis framework for real-time sound generation in interactive world models. By factorizing generation into chunk-wise autoregressive flow matching, \model synthesizes $48$\,kHz stereo audio in $100$\,ms chunks with an execution latency of $41.2$\,ms per chunk on a single NVIDIA H100 GPU ($\mathrm{RTF}=0.41$). Our decoupled visual conditioning couples long-term planning via a bounded $5$\,s Ring-KV cache with frame-aligned local guidance in the flow head, while training-only ShiftNCE enforces precise temporal synchronization with zero inference overhead. Evaluated on open-domain videos and interactive environments, \model achieves acoustic fidelity and spatial localization competitive with or exceeding offline bidirectional models, while seamlessly supporting real-time prompt updates.

\Needspace{15\baselineskip}
\section{Limitations}

\paragraph{Causal audio representations.}
\model generates audio in the continuous latent space of a frozen causal stereo VAE adapted from SoundReactor~\citep{soundreactor2025}. SoundReactor~\citep{soundreactor2025} reports lower reconstruction quality for its causal VAE decoder than for its non-causal counterpart. Developing more expressive, large-scale causal audio codecs will be crucial to close this performance gap in future work.

\paragraph{Stereo data scale.}
High-quality spatial audio data remains scarce compared to massive monophonic corpora. Many public stereo videos contain artificial or noisy channel separation that does not accurately reflect visual object motion. While our two-stage filtering pipeline and panoramic soundfield decoding extract reliable spatial supervision, expanding the scale of authentic stereo data remains a key goal for future work.

\bibliography{references}
\bibliographystyle{references}

\newpage
\appendix
\section{Implementation Configuration and Architectural Details}
\label{app:implementation}

Table~\ref{tab:implementation-config} summarizes the concrete configuration and architectural choices of \model, supplementing the high-level system description in Section~\ref{sec:method}.

\begin{table}[h]
\centering
\small
\caption{System implementation and architectural configuration of \model.}
\label{tab:implementation-config}
\begin{tabularx}{\linewidth}{>{\bfseries}l X}
\toprule
Component & Configuration and Architectural Specification \\
\midrule
Streaming Unit & $100$\,ms chunk: $3$ latent frames and $3$ video frames at $30$\,FPS \\
Causal Stereo VAE & Frozen SoundReactor~\citep{soundreactor2025} VAE; maps $48$\,kHz stereo audio to $30$\,Hz latents \\
AR Context Window & Ring-KV cache over $W = 50$ chunks ($5$\,s) \\
AR Backbone & $36$ layers, width $1{,}024$, $16$ attention heads, SwiGLU~\citep{shazeer2020glu} hidden size $4{,}096$ \\
Visual Encoder & Frozen DINOv3 S+~\citep{simeoni2025dinov3}; aspect-preserving resolution buckets; patch grid pooled $2\times$ \\
Visual Adapter & Linear bottleneck to $128$ channels, concatenated with $\Delta\mathbf{G}_n$; 2D-RoPE~\citep{su2024roformer,heo2024ropevit}; $1$ learned query per frame plus $1$ chunk-summary query \\
Text Prompt Encoder & Frozen T5Gemma~2~\citep{zhang2025t5gemma2}, truncated at $192$ tokens \\
Prompt Compressor & $1$ global token $+ 31$ learned queries, cached until the prompt updates \\
Flow Head & Width $1{,}024$, SwiGLU hidden size $5{,}120$; bidirectional attention within chunk; cross-attention to $3$ frame tokens; AdaLN-Zero~\citep{peebles2023dit} conditioned on $\mathbf{h}_t$ \\
Sampler & $15$ Euler steps per $100$\,ms chunk \\
Guidance & AR stage: $s_v = s_p = 3.0$. Flow head: AR-summary scale $1.0$, local-visual scale $8.0$ with APG~\citep{sadat2024apg} ($\eta=0$, momentum $+0.25$, radius $0.124$) \\
ShiftNCE & Training-only objective, using a frozen Synchformer~\citep{iashin2024synchformer} teacher \\
\bottomrule
\end{tabularx}
\end{table}

\paragraph{AR token pairing.}
At each step, the model receives the audio summary from the previous chunk and the visual summary from the current chunk. The audio summary is causally masked from attending to current or future visual tokens. Positional encoding uses chunk time indices via 2D-RoPE, maintaining continuous temporal representations when the Ring-KV buffer wraps.

\paragraph{Guidance formulations.}
\label{app:guidance}
Visual and textual controls are guided independently using classifier-free guidance~\citep{ho2022cfg}. Sampling evaluates a conditional branch $f$, a null-video branch $f_{\neg v}$, and a null-prompt branch $f_{\neg p}$:
\begin{equation}
	\hat f = f + (s_v - 1)(f - f_{\neg v}) + (s_p - 1)(f - f_{\neg p}).
	\label{eq:joint-cfg}
\end{equation}
Guidance scales are listed in Table~\ref{tab:implementation-config}. On the flow head, APG~\citep{sadat2024apg} is applied to the clean latent estimate $\hat{\mathbf{a}}_t = \mathbf{x}_{t,s} + (1-s)\mathbf{u}$, where $\mathbf{u}$ is the predicted velocity.

\paragraph{Baseline evaluations and clip windows.}
For fair comparison, AudioX~\citep{tian2025audiox}, ThinkSound~\citep{liu2025thinksound}, and PrismAudio~\citep{liu2026prismaudio} receive identical audio-centric captions as \model and are executed via their official inference pipelines with chain-of-thought reasoning disabled. V-AURA~\citep{viertola2025vaura} operates in video-only mode. On $5$\,s and $10$\,s benchmarks, all baseline models generate full-length clips in a single inference call. On the $30$\,s benchmark, AudioX and PrismAudio generate three contiguous $10$\,s clips matching their maximum context limits, whereas ThinkSound and V-AURA process the entire $30$\,s sequence in a single call. In contrast, \model continuously streams $100$\,ms chunks under bounded Ring-KV state with six sequential $5$\,s audio prompts. For $10$\,s caption-conditioned baseline runs, prompt pairs are concatenated in chronological order (e.g., ``First five seconds: \ldots\ Next five seconds: \ldots'').

\section{Training Recipe and Optimization Details}
\label{app:training-config}

Training consists of two distinct stages (Table~\ref{tab:stage-data}). Full optimization hyperparameters are listed in Table~\ref{tab:training-config}. The audio VAE~\citep{soundreactor2025}, DINOv3 S+~\citep{simeoni2025dinov3}, T5Gemma~2~\citep{zhang2025t5gemma2}, and the Synchformer~\citep{iashin2024synchformer} teacher remain frozen throughout training.

\begin{table}[h]
\centering
\small
\caption{Training curriculum: corpora, task mixtures, and view lengths.}
\label{tab:stage-data}
\begin{tabularx}{\textwidth}{p{0.18\linewidth} X p{0.18\linewidth} p{0.22\linewidth}}
\toprule
Stage & Corpus Composition & Task Ratio (video-audio${:}$audio-only) & View Representation \\
\midrule
Pretraining ($0$--$140$k) & Full video-audio/audio-only mixture: base stereo datasets plus mono-dominant expansion & $2{:}1$ & $5$\,s blocks, with $10$\,s paired views on 4 of 5 epochs \\
\addlinespace
Fine-tuning ($140$--$150$k) & High-synchrony filtered stereo video-audio only (expansion excluded) & $1{:}0$ & $5$\,s blocks, with $10$\,s paired views on 4 of 5 epochs \\
\bottomrule
\end{tabularx}
\end{table}

\begin{table}[h]
\centering
\small
\caption{Training configuration and hyperparameter settings.}
\label{tab:training-config}
\begin{tabularx}{\linewidth}{>{\bfseries}l X}
\toprule
Hyperparameter & Setting / Value \\
\midrule
Training Steps & $140{,}000$ pretraining steps $+ 10{,}000$ fine-tuning steps ($150{,}000$ total) \\
Optimizer & AdamW ($\beta_1 = 0.9$, $\beta_2 = 0.99$, weight decay $0.05$) \\
Learning Rate & Linear warmup from $0$ to $5 \times 10^{-5}$ over $4{,}000$ steps; constant $5 \times 10^{-5}$ thereafter \\
Gradient Clipping & Maximum gradient norm $1.0$ \\
EMA Rate & Decay rate $0.9999$ \\
Distributed Setup & $16$ NVIDIA H100 GPUs ($2$ nodes); batch size $16$ per GPU ($256$ global batch) \\
Flow Batch Multiplier & $4$ independent flow timesteps sampled per audio instance \\
Explorative Modeling & $K = 3$ candidate source noise vectors per example; backpropagation through best candidate \\
AR Causal Context & $5.0$\,s sliding window ($50$ chunks) under teacher forcing \\
Guidance Dropout & Video dropout $0.1$, prompt dropout $0.1$; flow-head categorical dropout $0.2$ ($0.1$ AR summary, $0.1$ local visual); most recent teacher-forced audio token nulled with $0.2$ probability \\
Loss Objectives & Joint Flow Matching ($\mathcal{L}_{\mathrm{flow}}$) and Causal ShiftNCE ($\mathcal{L}_{\mathrm{sync}}$) \\
\bottomrule
\end{tabularx}
\end{table}

\subsection{Explorative Modeling for Flow Matching (XM3)}
\label{app:explorative-fm}
During pretraining, we apply Explorative Modeling~\citep{gladstone2026explorative} to the flow-matching objective. For each training instance, we sample $K = 3$ noises with the same timestep $s \sim \mathcal{U}[0, 1]$ and dropout mask, then compute their vector field errors without gradients:
\begin{equation}
    \ell^{(k)} = \left\| \hat{\mathbf{u}}^{(k)} - (\mathbf{a}_t - \boldsymbol{\epsilon}^{(k)}) \right\|_2^2,
\end{equation}
and execute backpropagation solely on $k^* = \arg\min_k \ell^{(k)}$. Standard Gaussian noise sampling is used during inference.

\subsection{Ambiguity-Aware Causal ShiftNCE}
\label{app:shiftnce-details}
ShiftNCE distills synchronization signals from a frozen Synchformer~\citep{iashin2024synchformer} teacher into the AR backbone during training, and is disabled during inference.

\paragraph{Teacher representations.}
The teacher processes a causal $640$\,ms visual window ending on a chunk boundary, outputting eight ordered tokens $\mathbf{S}_t$. A 2-layer MLP maps the AR representation $\mathbf{h}_t$ to this alignment space. Pairwise similarity averages cosine similarity across the eight token slots:
\begin{equation}
    \mathrm{sim}(P(\mathbf{h}_i), \mathbf{S}_j) = \frac{1}{8}\sum_{k=1}^8 \cos\big(P(\mathbf{h}_i)_k, (\mathbf{S}_j)_k\big).
\end{equation}

\paragraph{Negative mining and anchor selection.}
For chunk $t$, candidate negatives comprise same-clip temporal offsets $\delta \in \{\pm 1, \pm 2, \pm 4\}$. Candidates exhibiting high teacher similarity ($\mathrm{sim}(\mathbf{S}_t, \mathbf{S}_{t+\delta}) \ge 0.97$) are filtered out to avoid false-negative penalization. Among valid sequence anchors, we select the top $40\%$ most teacher-separable anchors ranked by contrastive margin.

\paragraph{Objective formulation.}
With a fixed temperature $\tau = 0.07$, the alignment loss is averaged uniformly over selected valid anchors $\mathcal{A}_t$:
\begin{equation}
    \mathcal{L}_{\mathrm{sync}} = -\frac{1}{|\mathcal{A}_t|}\sum_{i \in \mathcal{A}_t} \log \frac{\exp\left( \mathrm{sim}(P(\mathbf{h}_i), \mathbf{S}_i) / \tau \right)}{\exp\left( \mathrm{sim}(P(\mathbf{h}_i), \mathbf{S}_i) / \tau \right) + \sum_{\delta \in \mathcal{D}_i} \exp\left( \mathrm{sim}(P(\mathbf{h}_i), \mathbf{S}_{i+\delta}) / \tau \right)}.
\end{equation}

\paragraph{Gradient balancing.}
The loss weight $\lambda_{\mathrm{sync}}$ dynamically maintains the synchronization gradient norm at approximately $5\%$ of the flow matching gradient norm, bounded within $[0.02, 0.08]$ after a $5{,}000$-step linear warmup. For audio-only clips, $\mathcal{L}_{\mathrm{sync}}$ is set to zero.

\subsection{Dynamic Prompt Schedules}
\label{app:prompt-schedules}
To facilitate mid-stream steering during continuous generation sessions, each 10-s paired training sequence is assigned one of four prompt schedules:
\begin{itemize}
    \item \textbf{Mid-stream switch ($60\%$)}: Prompt $P_1$ conditions the first 5-s segment, switching to prompt $P_2$ at the 5-s boundary.
    \item \textbf{Instruction withdrawal ($10\%$)}: Prompt $P_1$ conditions the first 5-s segment, followed by generation without text conditioning ($P = \emptyset$) for the second 5 seconds.
    \item \textbf{Instruction arrival ($10\%$)}: Generation begins without text conditioning ($P = \emptyset$) for the first 5 seconds, followed by prompt $P_1$ for the remaining 5 seconds.
    \item \textbf{Sustained hold ($20\%$)}: Prompt $P_1$ is maintained continuously across the full 10-s sequence.
\end{itemize}

\section{Training Corpus and Data Curation}
\label{app:data-accounting}

\subsection{Corpus Accounting and Composition}
The curated training dataset comprises $993{,}920$ total entries amounting to $1{,}464.93$ hours of audio. Table~\ref{tab:corpus-breakdown} presents the complete breakdown across all four data pools. An independent held-out validation set of $6{,}106$ entries is isolated from training.

\begin{table}[h]
\centering
\small
\caption{Training corpus breakdown by pool and audio duration.}
\label{tab:corpus-breakdown}
\begin{tabularx}{\linewidth}{X c c c c}
\toprule
Data Pool & 5-s Clips & 10-s Paired Views & Total Entries & Audio Duration (h) \\
\midrule
Base stereo video-audio & $590{,}510$ & $139{,}079$ & $590{,}510$ & $820.15$ \\
Mono-dominant video-audio & $128{,}885$ & $48{,}105$ & $128{,}885$ & $179.01$ \\
Base audio-only & $66{,}859$ & $60{,}832$ & $127{,}691$ & $261.84$ \\
Recovered audio-only & $146{,}834$ & -- & $146{,}834$ & $203.94$ \\
\midrule
Total Training Inventory & $933{,}088$ & $60{,}832$ & $993{,}920$ & $1{,}464.93$ \\
\bottomrule
\end{tabularx}
\end{table}

Video-audio $10$\,s paired views reuse existing $5$\,s clips and are not counted again in the entry or duration totals. The total row's $10$\,s column counts only the $60{,}832$ independent audio-only entries, which are included in both totals.

\paragraph{Source-family breakdown.}
The base stereo video-audio pool ($820.15$\,h) combines four sources:
\begin{itemize}
    \item \textit{Open-stereo in-the-wild} ($183{,}271$ clips / $254.54$\,h): VGGSound~\citep{chen2020vggsound} and AudioSet~\citep{gemmeke2017audioset};
    \item \textit{Interactive world model} ($135{,}478$ clips / $188.16$\,h): gameplay and simulated scenes;
    \item \textit{Panoramic ambisonics} ($151{,}298$ clips / $210.13$\,h): Sphere360~\citep{liu2025omniaudio} ($112{,}502$ clips / $156.25$\,h) and YT-AmbiGen~\citep{kim2025visage} ($38{,}796$ clips / $53.88$\,h); and
    \item \textit{Action-centric and egocentric video} ($120{,}463$ clips / $167.31$\,h): Kinetics-700~\citep{carreira2019kinetics700} ($92{,}474$ clips / $128.44$\,h) and HD-EPIC~\citep{perrett2025hdepic} ($27{,}989$ clips / $38.87$\,h).
\end{itemize}
The mono-dominant video-audio pool ($179.01$\,h) supplements events from VGGSound ($98{,}357$ clips / $136.61$\,h) and AudioSet ($30{,}528$ clips / $42.40$\,h). These clips retain two-channel audio, with ``mono-dominant'' referring to duplicated-mono or narrow-stereo signals with limited channel separation. This expansion is screened with Qwen3-Omni and used only during pretraining. It is excluded from spatial fine-tuning. Base audio-only data combines AudioCaps~\citep{kim2019audiocaps} and IF-Caps~\citep{tian2025audiox} ($127{,}691$ clips / $261.84$\,h). The recovered audio-only pool ($146{,}834$ clips / $203.94$\,h) retains weakly-aligned clips lacking direct visual spatial cues but containing clean acoustic events.

\subsection{Signal-Level Stereo Screening and Ambisonic Decoding}
\label{app:stereo-filter}
The following signal-level screening applies to the base stereo video-audio pool, rather than the separately retained, Qwen3-Omni-screened mono-dominant expansion:
\begin{enumerate}
    \item \textbf{Channel energy and DC centering}: Audio is resampled to $48$\,kHz and DC-centered per channel. Any clip exhibiting root-mean-square energy $\mathrm{RMS} < 10^{-4}$ on either channel is discarded as silent.
    \item \textbf{Lateral energy contrast and correlation}: We compute the signed energy contrast $d = (E_R - E_L)/(E_R + E_L)$ and normalized inter-channel cross-correlation $\rho \in [-1, 1]$. Clips with $|\rho| > 0.98$ or stereo width $< 0.15$ are excluded from the base stereo video-audio pool.
\end{enumerate}

\paragraph{Ambisonic-to-stereo rendering.}
For Sphere360 and YT-AmbiGen, the soundfield is rotated to align with the primary horizontal source and decoded at $\pm 45^\circ$, matching a $90^\circ$ perspective crop:
\begin{equation}
    L = W + \frac{\sqrt{2}}{2}(X + Y), \qquad R = W + \frac{\sqrt{2}}{2}(X - Y).
\end{equation}

\subsection{Multimodal Verification and Audio-Centric Captioning}
\label{app:captioning}
Qwen3-Omni~\citep{xu2025qwen3omni} scores audio-visual consistency from 1 to 5 and flags non-diegetic music. Video-audio clips below 4, or with such music, are removed. Captions name category, material, direction, reverberation, and envelope, and they omit visual appearance. Video-audio captions see the video and the audio, whereas audio-only captions see the audio only.

\subsection{Evaluation Set Isolation and Leakage Audit}
Evaluation covers five main held-out test splits: VGGSound ($5$\,s, $10$\,s) and Interactive Gameplay and Real-World ($5$\,s, $10$\,s with $4{,}096$ clips each; $30$\,s with $1{,}024$ clips). We additionally evaluate on Greatest Hits~\citep{owens2016visually} physical collisions ($244$ clips).

We audited the evaluation populations against our training inventory to verify strict data isolation:
\begin{itemize}
    \item On VGGSound, our evaluation adheres to the official test split, ensuring clip-level isolation against the training corpus.
    \item On Interactive benchmarks, source-group checks confirmed zero overlap with the training and rollout inventories. The $4{,}096$ clips in the 5-s and 10-s sets derive from $3{,}111$ unique source groups, with at most two non-overlapping windows per source. The $30$\,s benchmark contains $1{,}024$ distinct held-out sources.
\end{itemize}

\section{Evaluation Protocols and Metric Specifications}
\label{app:eval-protocols}

\subsection{Objective Metrics}
Metric $\mathrm{FD}_O$ computes the Fr\'echet Audio Distance directly on the mid-channel signal $(L+R)/2$, rather than averaging per-channel distances. Similarly, $\mathrm{S\text{-}FD}_O$ evaluates the side-channel signal $(L-R)/2$. DeSync is the mean absolute audiovisual synchronization offset, measured in seconds. Lower values indicate better synchronization. Segment-aligned CLAP score details and BiasSkill specifications are presented in Appendix~\ref{app:prompt-control} and Appendix~\ref{app:biasskill}, respectively. All other objective metrics utilize standard feature extractors as described in Section~\ref{sec:experiments}.

\subsection{Physical Onset Evaluation Protocol}
\label{app:onset-evaluation}
Greatest Hits~\citep{owens2016visually} onset settings follow MMAudio~\citep{mmaudio2024}: $244$ clips, $8$\,s, $22.05$\,kHz; threshold $\delta = 0.3$; $50$\,ms pre/post-peak context; $300$\,ms matching tolerance.

\subsection{Hardware Benchmark and Streaming Latency Protocol}
\label{app:latency-protocol}
Inference latency is measured on a single NVIDIA H100 GPU using batch size $1$ and $15$ Euler sampling steps per chunk. Steady-state synthesis per $100$\,ms chunk requires $41.21$\,ms at p50 and $41.38$\,ms at p95, achieving a Real-Time Factor (RTF) of $0.41$. Over a continuous $10$\,s streaming session including cold-start initialization, the overall RTF is $0.466$. During interactive prompt updates, chunk latency increases slightly to $45.28$\,ms at p50 and $47.40$\,ms at p95 due to state cache refresh.

\subsection{Baseline Reproduction and Access Mode Analysis}
\label{app:baselines}
We compare \model against state-of-the-art bidirectional stereo models (AudioX~\citep{tian2025audiox}, ThinkSound~\citep{liu2025thinksound}, PrismAudio~\citep{liu2026prismaudio}) and the streaming monophonic AR baseline V-AURA~\citep{viertola2025vaura}. All baselines are evaluated using official publicly released checkpoints, default samplers, and recommended classifier-free guidance parameters.

Baselines fall into three temporal access modes:
\begin{itemize}
    \item \textbf{Bidirectional access (AudioX, ThinkSound, PrismAudio)}: These models require full-clip visual context prior to inference. They process video frames via full-sequence bidirectional attention, rendering them unsuitable for streaming or live-deployment scenarios.
    \item \textbf{Stream mode (V-AURA)}: For long-form generation, V-AURA operates with $640$\,ms strides over a $2.56$\,s visual window. Within this localized window, frame encoding via TimeSformer~\citep{bertasius2021timesformer} remains non-causal, and DAC~\citep{kumar2023dac} neural decoding incorporates future context.
    \item \textbf{Causal mode (\model)}: Visual conditioning is limited to past and current chunk frames, maintaining strict causal alignment as detailed in Section~\ref{sec:method}.
\end{itemize}

\subsection{ITD-Based Metrics and Limitations}
\label{app:spatial-metrics}
In the production stereo used in our data, channel differences do not reliably reflect physical arrival-time delays. In our diagnostic experiment, duplicating a mono signal across both channels ($L = R$) achieved the lowest ITD-MSE~\citep{chen2022stereocrw} among the tested outputs despite having no stereo separation. We therefore omit ITD-MSE from our spatial evaluation. We also exclude FSAD~\citep{sun2024bewo} because it uses ITD-related StereoCRW features. Instead, we report $\mathrm{S\text{-}FD}_O$ and BiasSkill to evaluate stereo energy distribution and left/right channel agreement.

\section{Stereo Balance Agreement (BiasSkill)}
\label{app:biasskill}

Stereo balance agreement (BiasSkill) measures whether generated and reference audio favor the same left/right channel within $1$\,s windows. It normalizes the gain over a permutation baseline to account for a model's fixed channel bias.

\paragraph{Windowing and energy contrast.}
Audio signals are resampled to $24$\,kHz and segmented into non-overlapping $1$\,s windows. Each window is DC-centered per channel. For channel mean-square energies $E_L = \frac{1}{|W|}\sum_{n \in W} x_L[n]^2$ and $E_R = \frac{1}{|W|}\sum_{n \in W} x_R[n]^2$, the left--right energy contrast is:
\begin{equation}
    d = \frac{E_R - E_L}{E_R + E_L + 10^{-12}}.
\end{equation}
A window is designated right-dominant if $d > 0$ and left-dominant if $d < 0$, with $|d|$ reflecting the magnitude of directional separation.

\paragraph{Gating and eligibility criteria.}
To prevent scoring ambiguous or silent intervals, a ground-truth window is considered eligible only if it satisfies:
\begin{equation}
    |d_{\mathrm{GT}}| \ge 0.3, \qquad \mathrm{RMS}_{\mathrm{GT}} = \sqrt{\frac{E_L + E_R}{2}} \ge 10^{-4}.
\end{equation}
For generated audio, we enforce an absolute activity threshold of $\mathrm{RMS}_{\mathrm{gen}} \ge 0.001$ ($-60$\,dBFS). A generated window is marked as correct if and only if:
\begin{equation}
    |d_{\mathrm{gen}}| \ge 0.1 \quad \text{and} \quad d_{\mathrm{gen}} \cdot d_{\mathrm{GT}} > 0.
\end{equation}
Critically, near-mono generations ($|d_{\mathrm{gen}}| < 0.1$) that fail to establish stereo separation are counted as incorrect (retained in the evaluation denominator), preventing models from achieving artificially inflated accuracy by producing centered outputs.

\paragraph{Balanced accuracy.}
Accuracy is computed over eligible windows with ground-truth side $s \in \{L, R\}$ in clip $i$, macro-averaged across clips $C_s$, and balanced between channels:
\begin{equation}
    A_s = \frac{1}{|C_s|} \sum_{i \in C_s} \frac{1}{|W_{i,s}|} \sum_{w \in W_{i,s}} \mathbb{1}[\hat{y}_{i,w} = s], \qquad A_{\mathrm{bal}} = \frac{1}{2}(A_L + A_R).
\end{equation}
Table~\ref{tab:biasskill-coverage} summarizes the eligible window counts and retention rates across test sets.

\begin{table}[h]
\centering
\small
\caption{Eligible $1$\,s window counts and retention rates for BiasSkill evaluation.}
\label{tab:biasskill-coverage}
\setlength{\tabcolsep}{8pt}
\begin{tabular}{@{}lccccc@{}}
\toprule
Set & Clips & Cand.\ Win. & Elig.\ Clips & Elig.\ Win.\ (L / R) & Keep (\%) \\
\midrule
VGGSound $10$\,s & $4{,}096$ & $40{,}960$ & $1{,}589$ & $3{,}314$ / $3{,}356$ & $16.3$ \\
Interactive $10$\,s & $4{,}096$ & $40{,}960$ & $2{,}275$ & $6{,}759$ / $3{,}316$ & $24.6$ \\
Interactive $30$\,s & $1{,}024$ & $30{,}720$ & $679$ & $2{,}870$ / $3{,}200$ & $19.8$ \\
\midrule
Total & $9{,}216$ & $112{,}640$ & -- & $12{,}943$ / $9{,}872$ & $20.3$ \\
\bottomrule
\end{tabular}
\end{table}

\paragraph{Permutation null and normalized BiasSkill.}
\begin{table}[t]
\centering
\caption{Detailed stereo balance agreement results under the $-60$\,dBFS generation-RMS gate. L/R Rec.\ denotes per-side recall; Bal.\ Acc.\ is $A_{\mathrm{bal}}$; RMS Cov.\ is the coverage of $\mathrm{RMS}_{\mathrm{gen}} \ge 0.001$; Act.\ Cov.\ requires both RMS and $|d_{\mathrm{gen}}| \ge 0.1$; $\Delta = A_{\mathrm{bal}} - \mu_{\mathrm{null}}$; BiasSkill represents the normalized effect size (\%); and $p$ indicates the one-sided permutation test value.}
\label{tab:biasskill-full}
\scriptsize
\setlength{\tabcolsep}{3pt}
\resizebox{\linewidth}{!}{%
\begin{tabular}{@{}llccccccccc@{}}
\toprule
Set & Model & L Rec. & R Rec. & Bal.\ Acc. & RMS Cov. & Act.\ Cov. & Null $\mu_{\mathrm{null}}$ & $\Delta$ & BiasSkill (\%) & $p$ \\
\midrule
\multirow{4}{*}{VGG $10$\,s}
 & AudioX & 0.273 & 0.186 & 0.229 & 0.990 & 0.430 & 0.199 & $+0.030$ & 3.79 & 0.0002 \\
 & ThinkSound & 0.181 & 0.386 & 0.283 & 0.956 & 0.554 & 0.260 & $+0.024$ & 3.22 & 0.0035 \\
 & PrismAudio & 0.277 & 0.093 & 0.185 & 0.946 & 0.364 & 0.176 & $+0.009$ & 1.10 & 0.1052 \\
\rowcolor{modelrow} & \modelours & 0.256 & 0.211 & 0.233 & 0.976 & 0.439 & 0.198 & $+0.035$ & \textbf{4.42} & 0.0001 \\
\midrule
\multirow{4}{*}{Int.\ $10$\,s}
 & AudioX & 0.256 & 0.237 & 0.246 & 0.967 & 0.460 & 0.222 & $+0.024$ & 3.08 & 0.0001 \\
 & ThinkSound & 0.159 & 0.395 & 0.277 & 0.983 & 0.594 & 0.290 & $-0.013$ & $-1.78$ & 0.9568 \\
 & PrismAudio & 0.387 & 0.036 & 0.211 & 0.995 & 0.447 & 0.212 & $-0.001$ & $-0.06$ & 0.5390 \\
\rowcolor{modelrow} & \modelours & 0.317 & 0.273 & 0.295 & 0.995 & 0.559 & 0.250 & $+0.045$ & \textbf{6.03} & 0.0001 \\
\midrule
\multirow{4}{*}{Int.\ $30$\,s}
 & AudioX & 0.247 & 0.225 & 0.236 & 0.999 & 0.468 & 0.236 & $0.000$ & $-0.01$ & 0.4974 \\
 & ThinkSound & 0.101 & 0.455 & 0.278 & 0.997 & 0.559 & 0.273 & $+0.005$ & 0.63 & 0.3088 \\
 & PrismAudio & 0.493 & 0.046 & 0.270 & 0.997 & 0.519 & 0.238 & $+0.032$ & 4.19 & 0.0004 \\
\rowcolor{modelrow} & \modelours & 0.447 & 0.397 & 0.422 & 0.995 & 0.716 & 0.313 & $+0.109$ & \textbf{15.90} & 0.0001 \\
\bottomrule
\end{tabular}%
}
\end{table}
Generative models often exhibit fixed channel biases stemming from training data or architecture (e.g., ThinkSound consistently generates right-heavy energy, whereas PrismAudio is left-heavy). To eliminate these baseline preferences, we draw $10{,}000$ whole-clip derangements shared across all models, pairing generated audio clips with mismatched ground-truth videos while maintaining internal window alignment.
The permutation null score $\mu_{\mathrm{null}}$ represents the expected balanced accuracy under these shuffled pairings. BiasSkill is defined as the normalized gain over this model-specific null:
\begin{equation}
    \mathrm{BiasSkill} = \frac{A_{\mathrm{bal}} - \mu_{\mathrm{null}}}{1 - \mu_{\mathrm{null}}} \times 100\%.
\end{equation}
A BiasSkill score of $0\%$ indicates performance no better than random chance under the model's fixed channel bias, while $100\%$ represents perfect left/right channel agreement on eligible windows. The one-sided Monte Carlo $p$-value is given by:
\begin{equation}
    p = \frac{1 + \sum_{b=1}^{B} \mathbb{1}[A_{\mathrm{bal}}^{(b)} \ge A_{\mathrm{bal}}]}{B + 1},
\end{equation}
with $B = 10{,}000$. Table~\ref{tab:biasskill-full} reports the detailed stereo balance agreement results.

\section{Interactive Text Control and Prompt-Switch Analysis}
\label{app:prompt-control}

\subsection{CLAP Implementation and Segment Alignment}
Semantic audio-text correspondence is evaluated using LAION-CLAP~\citep{wu2023clap}. Audio tracks are converted to mono at a $48$\,kHz sampling rate. Each waveform is divided into consecutive, non-overlapping $5$\,s segments and evaluated against its corresponding audio caption. Embeddings are extracted in FP16 precision and normalized in FP32. We report raw cosine similarities, macro-averaged within individual clips and across the test evaluation set.

\subsection{Paired Counterfactual Intervention Protocol}
To rigorously verify that dynamic acoustic changes during streaming stem from active prompt steering rather than underlying visual transitions in the video, we construct a paired counterfactual intervention across all $4{,}096$ clips in VGGSound $10$\,s and Interactive $10$\,s.

For each video clip with consecutive reference prompts $(P_1, P_2)$, we synthesize two paired generations sharing identical model parameters ($150$\,k EMA checkpoint), visual feature inputs, random seeds, Euler solver settings ($15$ steps), and classifier-free guidance scales:
\begin{itemize}
    \item \textbf{Switch Arm}: Conditioned on prompt $P_1$ for $0$--$5$\,s, dynamically switching to prompt $P_2$ at the $5$\,s boundary without clearing model state cache.
    \item \textbf{Hold Arm}: Conditioned continuously on prompt $P_1$ throughout the full $0$--$10$\,s duration.
\end{itemize}

Let $A_1$ and $A_2$ denote the generated $5$\,s audio segments, and let $s_{ij} = \cos(e_A(A_i), e_T(P_j))$ represent the pairwise cosine similarity between audio segment $A_i$ and text prompt $P_j$.
Segment alignment $S$, matching advantage $M$, and dual relative match indicator $R$ are formulated as:
\begin{align}
    S &= \frac{1}{2}(s_{11} + s_{22}), \\
    M &= \frac{1}{2}\big[(s_{11} - s_{12}) + (s_{22} - s_{21})\big], \\
    R &= \mathbb{1}[s_{11} > s_{12} \land s_{22} > s_{21}].
\end{align}
The second segment's preference for the updated prompt is given by $\Delta = s_{22} - s_{21}$. The net paired intervention gain $G_i$ for clip $i$ is defined as:
\begin{equation}
    G_i = \Delta_i^{\mathrm{switch}} - \Delta_i^{\mathrm{hold}}.
    \label{eq:prompt-switch-gain}
\end{equation}
Table~\ref{tab:prompt-control-full} summarizes the full paired intervention experimental results.

\begin{table}[h]
\centering
\small
\caption{Full paired-control intervention results ($4{,}096$ clips per set). The $s_{ij}$ entries are raw CLAP cosine similarities. $S$ is the mean of the aligned similarities $s_{11}$ and $s_{22}$, $M$ is the mean matching advantage, and $R$ is reported as a percentage (\%). Hold is evaluated against the same reference prompt pair $(P_1, P_2)$ as Switch.}
\label{tab:prompt-control-full}
\setlength{\tabcolsep}{5pt}
\begin{tabular}{@{}llccccccc@{}}
\toprule
Set & Arm & $s_{11}$ & $s_{12}$ & $s_{21}$ & $s_{22}$ & $S$ & $M$ & $R$ (\%) \\
\midrule
\multirow{2}{*}{VGG $10$\,s}
 & Switch & 0.3823 & 0.3312 & 0.3293 & 0.3530 & 0.3677 & 0.0374 & 25.68 \\
 & Hold   & 0.3823 & 0.3312 & 0.3543 & 0.3270 & 0.3546 & 0.0119 & 14.43 \\
\midrule
\multirow{2}{*}{Int.\ $10$\,s}
 & Switch & 0.3560 & 0.3062 & 0.3062 & 0.3284 & 0.3422 & 0.0360 & 23.44 \\
 & Hold   & 0.3560 & 0.3062 & 0.3316 & 0.3013 & 0.3286 & 0.0097 & 12.82 \\
\bottomrule
\end{tabular}
\end{table}

\subsection{Confidence Intervals and Prompt Stratification}
Confidence intervals for the mean intervention gain $G$ are derived via $20{,}000$ paired bootstrap resamples of clips within each benchmark dataset (random seed $5031$). We report the $2.5\text{th}$ and $97.5\text{th}$ percentiles as $[l, u]$. The compact $\pm$ margin corresponds to half of this interval width.

To evaluate how semantic contrast influences prompt control responsiveness, clips are ranked by text-to-text cosine similarity $\cos(e_T(P_1), e_T(P_2))$ and partitioned into four equal quartiles of $1{,}024$ clips each (Table~\ref{tab:prompt-control-strata}). Quartile Q1 contains the most semantically distinct prompt pairs, while Q4 contains the most similar. As shown, $G$ remains positive across all strata and scales directly with the degree of prompt divergence.

\begin{table}[h]
\centering
\small
\caption{Stratification of paired prompt gain $G$ across text-similarity quartiles ($1{,}024$ clips per quartile). Brackets denote $95\%$ clip-level bootstrap percentile intervals.}
\label{tab:prompt-control-strata}
\setlength{\tabcolsep}{7pt}
\begin{tabular}{@{}lccc@{}}
\toprule
Evaluation Set & Quartile & Text Cosine Range & Intervention Gain $G$ \\
\midrule
\multirow{4}{*}{VGGSound $10$\,s}
 & Q1 (Most Distinct) & $[-0.244, 0.525]$ & $0.1393$ $[0.1287, 0.1499]$ \\
 & Q2                 & $[0.525, 0.727]$  & $0.0441$ $[0.0388, 0.0494]$ \\
 & Q3                 & $[0.727, 0.842]$  & $0.0157$ $[0.0127, 0.0187]$ \\
 & Q4 (Most Similar)  & $[0.842, 0.990]$  & $0.0047$ $[0.0030, 0.0064]$ \\
\midrule
\multirow{4}{*}{Interactive $10$\,s}
 & Q1 (Most Distinct) & $[-0.287, 0.490]$ & $0.1277$ $[0.1187, 0.1367]$ \\
 & Q2                 & $[0.491, 0.688]$  & $0.0514$ $[0.0459, 0.0569]$ \\
 & Q3                 & $[0.688, 0.820]$  & $0.0247$ $[0.0210, 0.0284]$ \\
 & Q4 (Most Similar)  & $[0.820, 0.990]$  & $0.0063$ $[0.0046, 0.0080]$ \\
\bottomrule
\end{tabular}
\end{table}

\section{Subjective User Study Protocol}
\label{app:user-study}

The user study evaluates $40$ clips from Interactive $10$\,s, comprising $20$ gameplay and $20$ real-world scenes. Participants evaluate randomized, anonymized A/B pairs (with a tie option) across four dimensions: spatial alignment, temporal synchronization, semantic consistency, and overall preference. The five evaluated audio sources comprise Ground Truth, \model, AudioX, ThinkSound, and PrismAudio (V-AURA is excluded due to its monophonic output). Twenty independent evaluators produced a total of $400$ pairwise evaluations, yielding $40$ ratings per paired comparison.

\begin{figure}[p]
\centering
\includegraphics[width=\linewidth]{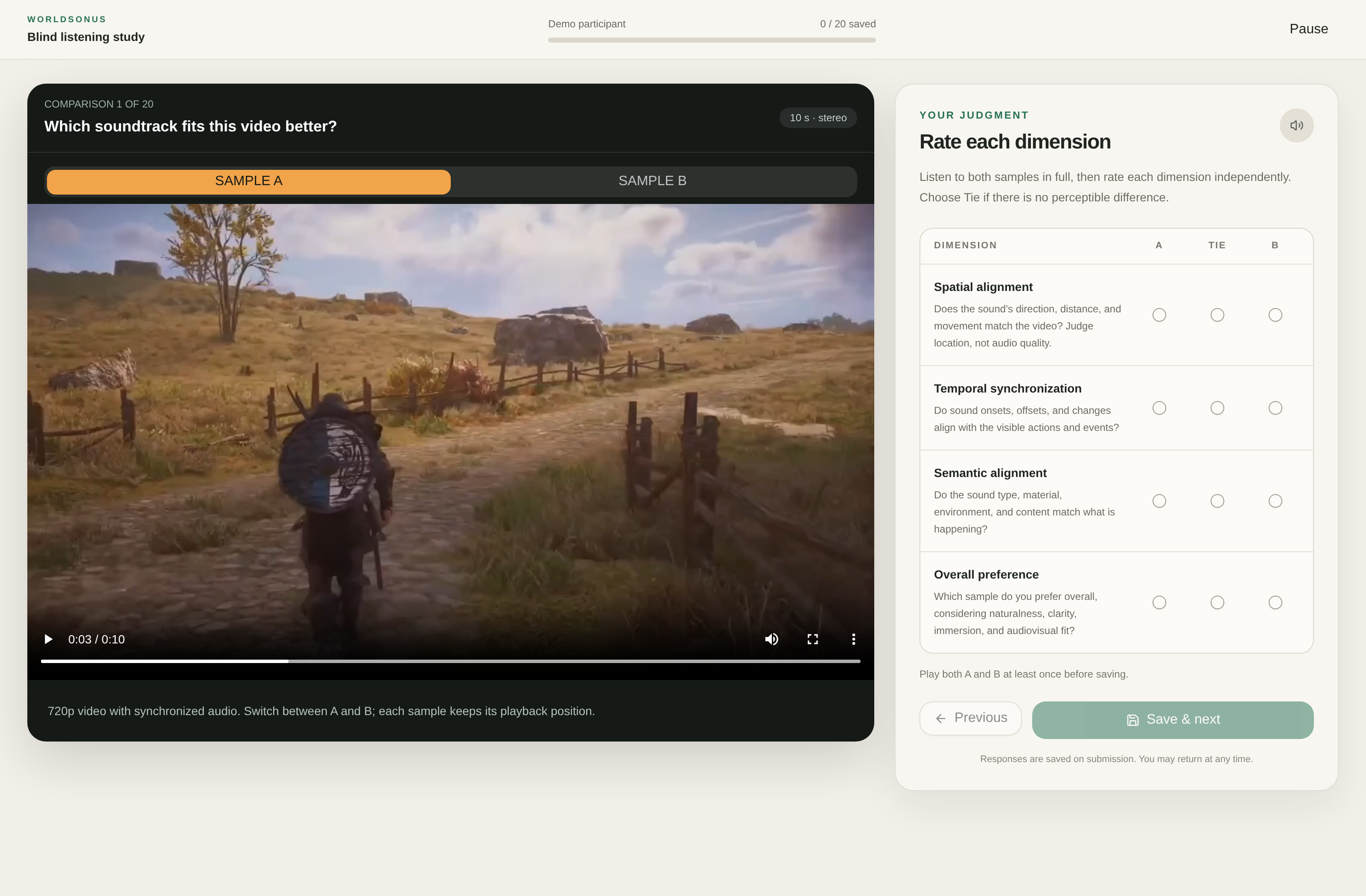}
\par\smallskip
{\small (a) Gameplay example}
\par\medskip
\includegraphics[width=\linewidth]{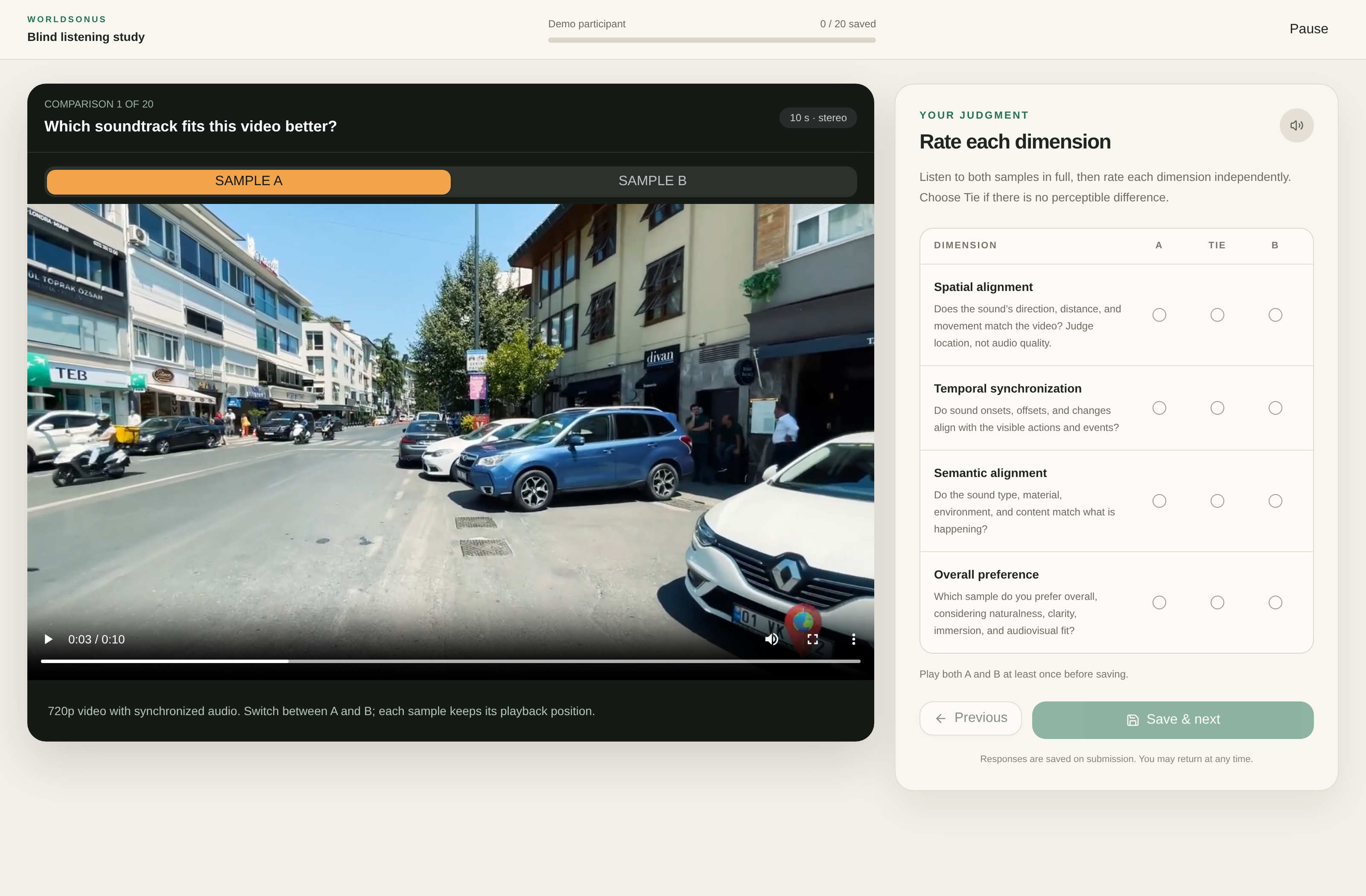}
\par\smallskip
{\small (b) Real-world example}
\caption{User-study interface for gameplay and real-world clips. Participants compare samples A and B and select A, Tie, or B for spatial alignment, temporal synchronization, semantic alignment, and overall preference.}
\label{fig:user-study-interface}
\end{figure}

Figure~\ref{fig:user-study-interface} shows the evaluation interface for gameplay and real-world clips, with paired samples labeled A and B and a separate A/Tie/B choice for each criterion.

The preference score $P_{ij}^{(d)}$ of method $i$ over method $j$ on criterion $d$ is calculated as:
\begin{equation}
    P_{ij}^{(d)} = 100 \times \frac{W_{ij}^{(d)} + \frac{1}{2}T_{ij}^{(d)}}{W_{ij}^{(d)} + T_{ij}^{(d)} + L_{ij}^{(d)}},
\end{equation}
where $W_{ij}^{(d)}$, $T_{ij}^{(d)}$, and $L_{ij}^{(d)}$ denote total wins, ties, and losses, respectively. Ties contribute half a vote ($0.5$), ensuring symmetric scoring $P_{ij}^{(d)} + P_{ji}^{(d)} = 100\%$. All evaluation matrices are reported on a normalized $0$--$100\%$ scale.

\end{document}